\documentclass[12pt]{article}
\ifx\pdfoutput\undefined
\usepackage[dvips,bookmarks]{hyperref}
\else
\usepackage{hyperref}
\fi
\hypersetup{colorlinks,bookmarksopen,bookmarksnumbered,citecolor=blue,
    pdfstartview=FitH}
\def\hhref#1{\href{http://arxiv.org/abs/hep-th/#1}{hep-th/#1}}
\def\mhref#1{\href{mailto:#1}{#1}}
\usepackage{amssymb,amsfonts,amsmath,color,graphicx}

\def\mysection#1{\section{#1}
    \setcounter{equation}{0}
    \renewcommand{\theequation}{\thesection.\arabic{equation}}}
\def\mysubsection#1{\subsection{#1}
    \setcounter{equation}{0}
    \renewcommand{\theequation}{\thesubsection.\arabic{equation}}}

\def\on#1#2{{\buildrel{\mkern2.5mu#1\mkern-2.5mu}\over{#2}}}
\def\f#1#2{{\textstyle{#1\over#2}}}    % fraction

\begin{document}

\begin{center}
{June 23, 2005 \hfill YITP-SB-05-15} \vskip.5in
{\bf{\huge\color{cyan} Conquest of the ghost\\[.15in]
pyramid of the superstring}}\\[.3in]
Kiyoung Lee and Warren Siegel\footnote{
\it \mhref{klee@insti.physics.sunysb.edu} {\rm and}
\mhref{siegel@insti.physics.sunysb.edu}}\\[.1in]
{\it C. N. Yang Institute for Theoretical Physics \\
State University of New York, Stony Brook, NY 11794-3840}\\[.5in]
\end{center}

\begin{abstract}
We give a new Becchi-Rouet-Stora-Tyutin operator for the superstring.
It implies a quadratic gauge-fixed action, and a new gauge-invariant action with first-class constraints.
The infinite pyramid of spinor ghosts
appears in a simple way through ghost gamma matrices.
\end{abstract}
\thispagestyle{empty}           % no heading or foot on first page (LaTeX)
\newpage

\mysection{Introduction}

The advantages of supersymmetry are somewhat obscured in the
Ramond-Neveu-Schwarz formalism, as is the case for supersymmetric
particle theories when not formulated in superspace.  For example,
cancellations of divergences are not obvious, and amplitudes with
many fermions are difficult to calculate.

Some of these problems were resolved with the Green-Schwarz
formalism, but it proved difficult to quantize except in the
lightcone gauge, where some manifest supersymmetry is retained in
trade for the loss of some manifest Lorentz invariance.  (Similar
remarks apply to the Casalbuoni-Brink-Schwarz superparticle.)  For
example, higher-point diagrams of any type are difficult to
calculate because longitudinal polarizations and momenta introduce
nonlinearities, and in particular cancellation of anomalies (or
any $\epsilon$-tensor contribution) is difficult to check.

Covariant quantization of the Green-Schwarz action was attempted
\cite{StonyBrook}.  A class of derivative gauges was introduced
that led to a pyramid of ghosts.  Counting arguments showed that
the conformal anomaly canceled, and summation of ghost
determinants agreed with the lightcone result due to the
``identity" $1-2+3-...=1/4$.  Unfortunately, due to a
noninvertible transformation the gauge-fixed action found by this
method proved not to be invariant under the
Becchi-Rouet-Stora-Tyutin transformations derived by the same
method \cite{toobad}.  This problem already appeared for the
Casalbuoni-Brink-Schwarz superparticle.

In the meantime, an alternative approach to the quantum
superparticle was developed \cite{freeSBRST}, based on adding
extra dimensions to the lightcone, a method that had successfully
given free gauge-invariant actions for arbitrary representations
of the Poincar\'e group in arbitrary dimensions \cite{Zwiebach}.
This approach directly gave a BRST operator with the right
cohomology.  Using the relation between this BRST operator and
Zinn-Justin-Batalin-Vilkovisky first-quantization \cite{ZJBV}, a
manifestly supersymmetric classical mechanics action for this
superparticle followed, including a BRST-invariant gauge-fixed
action \cite{ilk}.
A crucial difference from the previous method was that ``nonminimal" fields were required:  There was necessarily a ``pyramid" of ghosts, not just a linear tower.
However, because of a required Fierz identity,
the method of adding extra dimensions could not be directly
applied to the lightcone Green-Schwarz superstring.

Various alternatives for a manifestly supersymmetric superstring
have since been tried; the most successful is the pure spinor
formalism \cite{pure}.  It has proven somewhat more useful than
RNS or lightcone GS approaches in calculating tree amplitudes
\cite{puretree}; its application to loop amplitudes is in progress
\cite{pureloop}.  If the formalism for all loops is developed, it
should provide a simpler proof of finiteness, which previously
required a combination of RNS and lightcone GS results (and
equivalence of the two approaches).  The pure spinor approach has
two main shortcomings:

The first problem is the lack of a manifestly supersymmetric (and
Lorentz covariant) path-integration measure.  This is a problem in
all known superspace approaches to first-quantizing superparticles
and superstrings.  One consequence is that Green functions (or the
effective action in the superparticle case) are not manifestly
supersymmetric off shell.  Another is that gauge fixing the string
field theory (with ghost fields) is not simple.  We will not
address this problem here.

The other problem is that the pure-spinor BRST operator lacks the
$c$ and $b$ ghosts associated with the usual 2D coordinate
invariances (and their associated Virasoro constraints).  This is
directly related to the lack of a corresponding action with
worldsheet metric; the action is known only in the conformal
gauge.  Furthermore, the moduli that are the remnants of the
metric in the conformal gauge must be inserted by hand.  Another
consequence of the lack of these ghosts as fundamental variables
is that they must be reconstructed as complicated composite
operators for use as insertions in loop diagrams.  The
(gauge-fixed) action, BRST operator, moduli, and operator
insertions are thus separate postulates of the formalism, rather
than all following from a gauge-invariant action as in other
formalisms.

In this paper we will formulate the superstring with the
ghost structure indicated by the original attempt of
\cite{StonyBrook} and the successful treatment of the
superparticle in \cite{freeSBRST}:  the
usual $c$ and $b$ ghosts, and a pyramid of spinors
labeled by ghost number and generation.
The main result is the BRST operator (from which the gauge-invariant action follows), which takes the form
\begin{equation}
Q_{sstring}=U\left(\ \int
c~T~+~\f14\bar{\pi}\tilde{\gamma}^{\oplus}\pi|_{>}~\right)U^{-1}
\end{equation}
with
\begin{equation}
U=e^{\int\tilde{\theta}D}~e^{i\int R^{a}|_{>}P^{(\pm)}_{a}}~e^{\int(R^{\oplus}+\theta\tilde{\gamma}^{\oplus}\theta/2)|_{>}b}
\end{equation}
where $T$ is essentially the energy-momentum tensor, $D$ and $P$
are the usual ``covariant derivatives" in the affine Lie algebra
of the classical superstring, $\tilde\theta$ is a certain linear
combination of ghost $\theta$'s, $R^i$ are certain expressions
quadratic in $\theta$'s, $\pi$ is conjugate to $\theta$,
$\tilde\gamma^\oplus$ is a ghost partner to the gamma matrices
$\gamma^a$ (which act only on $\theta$ and $\pi$), and $|_>$ picks
out the ghost contributions.
The gauge-fixed Hamiltonian is just $\{Q,\int b\}=\int T$.
The unitary transformations are
necessary because they change the Hilbert space, and so cannot be
dropped:  A simple analog is the BRST operator for the spinning
(Dirac) particle in an external gauge field:
$$
Q_{Dirac} = e^{c\gamma^a\nabla_a/\gamma^\oplus} (\gamma^{\oplus 2} b)
e^{-c\gamma^a\nabla_a/\gamma^\oplus}
= \gamma^{\oplus 2} b + \gamma^\oplus\gamma^a\nabla_a -\f12 c(\gamma^a\nabla_a)^2
$$
where $1/\gamma^\oplus$ doesn't exist on the correct Hilbert space, but cancels when the ``unitary" transformation is evaluated.

We begin in section 2 by reviewing
the free superparticle, which has already been quantized
(and its BRST cohomology checked) in this approach.
Because of the similarity of the algebra of super
Yang-Mills \cite{algebra} to that of the superstring \cite{affine},
in section 3 we couple this superparticle
to external super Yang-Mills superfields.
We use an almost identical method
to derive the BRST operator for the superstring in section 4.
We finish with our conclusions in section 5.
(Mathematical details are relegated to the Appendices.)

\mysection{Review of free superparticle}

We will start from the free super BRST operator derived in
\cite{freeSBRST}.
The generic BRST operator for arbitrary fields (massless, or massive by dimensional reduction) is constructed by starting with a representation of the lightcone SO(D$-$2) (which defines the theory) and adding 4 bosonic and 4 fermionic dimensions to obtain a covariant representation, including all auxiliary fields and ghosts.  (This is somewhat redundant for bosons, but necessary for fermions.)
The resulting generators $S^{AB}$ of OSp(D,2$|$4) spin carry
 vector indices $A,B$ that are separated into the usual SO(D$-$1,1) indices $a,b$ and the rest as
\begin{equation}
\label{indices} A~=~(+,-,a;\mu,\tilde\mu) =
(+,-,i),~~~~\mu~=~(\oplus,\ominus)
\end{equation}
where $+,-$ belong to an SO(1,1) subgroup and $\mu,\tilde\mu$ to
two Sp(2)'s, of which only the diagonal subgroup will be useful.
The BRST operator then takes the generic form
\begin{equation}
\label{classicalbrst} Q'_{free}~=~\f{1}{2} c~\square
~+~S^{\oplus
a}\partial_{a}~+~S^{\oplus \oplus} b ~+~S^{\tilde\oplus -}~~~~~~~~~(~\square=\partial^{a}\partial_{a}~)\\
\end{equation}

In the case of the superparticle, the spin operators are
\begin{equation}
\label{sdef}S^{A B}~=~-\f{1}{4}~\bar{\eta}\Gamma^{[ A}\Gamma^{B \}}\eta\\
\end{equation}
in terms of self-conjugate variables $\eta$, which arose from the usual self-conjugate SO(D$-$2) fermionic spinor of lightcone superspace.
We decompose the OSp(D,2$|$4) gamma-matrices $\Gamma^A$ in terms of those of the subgroup SO(1,1) and those ($\gamma$) of the subgroup OSp(D$-$1,1$|$4) as
\begin{equation} \label{gamma}
\Gamma^i~=~\begin{pmatrix}
\gamma^i & 0 \\
0&-\gamma^i\end{pmatrix},~~~ \Gamma^{+}~=~\begin{pmatrix}
0 & -I \\
0&0\end{pmatrix},~~~\Gamma^{-}~=~\begin{pmatrix}
0 & 0 \\
-I&0\end{pmatrix}
\end{equation}
~~~~~~with (anti)commutation relations
\begin{eqnarray}
&&\{\gamma^{a},\gamma^{b}\}~=~-2\eta^{ab},~~~~~~~~\eta^{ab}=(-+++\cdots)\nonumber\\
&&\{\gamma^{a},\gamma^{\mu}\}~=~\{\gamma^{a},\tilde{\gamma}^{\mu}\}~=~0 \\
&&[\gamma^{\mu},\gamma^{\nu}]~=~[\tilde{\gamma}^{\mu},\tilde{\gamma}^{\nu}]~=~2C^{\mu\nu},
~~~[\gamma^{\mu},\tilde{\gamma}^{\nu}]~=~0
\end{eqnarray}
where $C^{\mu\nu}$ is the Sp(2) metric with convention
\begin{equation}
C^{\oplus\ominus}~=~C_{\ominus\oplus}~=~i~=~-C^{\ominus\oplus}~=~-C_{\oplus\ominus}
\end{equation}
and we have denoted $\gamma^{\tilde\mu}\equiv\tilde\gamma^\mu$ for
legibility. The generalization of the fermionic superspace
coordinate $\theta$ and its conjugate momentum appear through the
analogous decomposition
\begin{equation}
\label{qdef} \eta~=~\begin{pmatrix} \pi \\ \theta
\end{pmatrix},~~~~~~~~\pi~=~\frac{\partial}{\partial\theta}
\end{equation}
We begin with a chiral (Weyl) spinor $\eta$, and multiplication by any $\Gamma$ changes the chirality:  not just $\Gamma^a$ ($\gamma^a$) as usual, but also $\Gamma^\pm$, which shows that $\pi$ and $\theta$ have opposite chirality (as expected, since they are conjugate), and $\Gamma^\mu$ ($\gamma^\mu$).

\begin{figure}[h]
 \centering
\includegraphics[scale=.7]{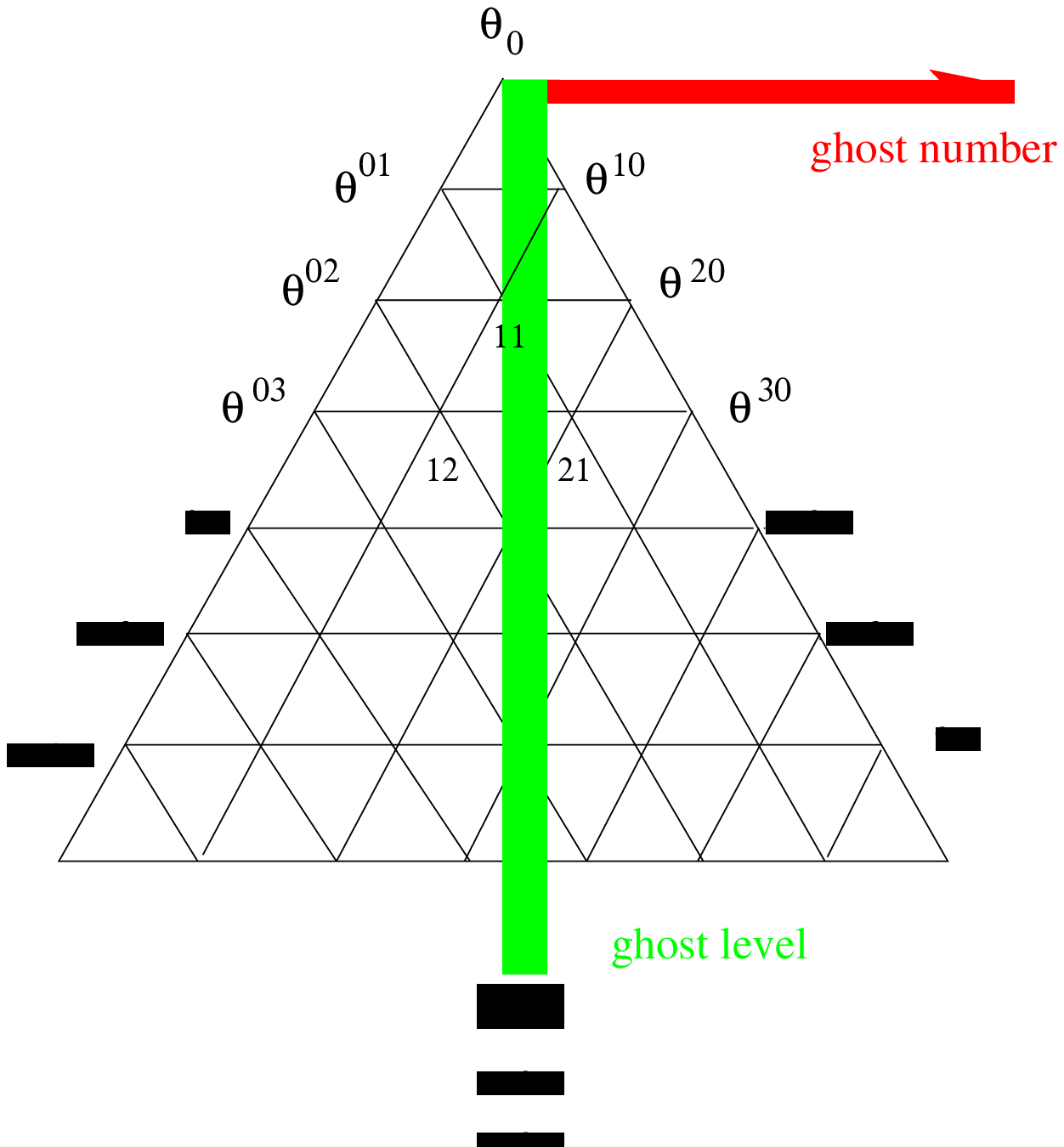}
 \caption{infinite pyramid of ghosts}
 \label{pyramid}
\end{figure}

 This BRST operator is supersymmetric and also has an infinite
pyramid of ghosts. To see these ghosts we need to define creation
and annihilation operators from $\gamma^\mu$ and
$\tilde{\gamma}^{\mu}$ as follows:
\begin{eqnarray}
\label{creatannihil} \gamma^{\mu}~=~a^{\mu}~+~a^{\dagger
\mu} &,& \tilde{\gamma}^{\mu}~=~i(a^{\mu}~-~a^{\dagger \mu})\\
~[a^{\mu},a^{\dagger \nu}]~&=&~C^{\mu\nu}\end{eqnarray} Then
$\theta$ can be expanded giving the usual physical supersymmetry
fermionic coordinate $\theta_{0}$ at the top of the infinite pyramid
of ghosts:

\begin{eqnarray}
\label{thetadef}
|{}^{p,q}\rangle&\equiv&i^{\f{(p+q)(p+q+1)}{2}}\frac{1}{\sqrt{p!}\sqrt{q!}}(a^{\dag\oplus})^{p}(a^{\dag\ominus})^{q}~|0\rangle\nonumber\\
\langle {}_{p,q}|&\equiv&(-i)^{\f{(p+q)(p+q+1)}{2}}\frac{1}{\sqrt{p!}\sqrt{q!}}\langle0|~(a_{\oplus})^{p}(a_{\ominus})^{q}\nonumber\\
\langle {}_{p,q}|{}^{r,s}\rangle&=&\delta_p^r\delta_q^s\nonumber\\
\theta^{p,q}&\equiv&\langle\theta|{}^{p,q}\rangle
\ = \ \theta^{p,q\dagger}\nonumber\\
\pi_{p,q}&\equiv&\langle {}_{p,q}|\pi\rangle
\end{eqnarray}
 where
\begin{equation}
\theta_{0}~\equiv~\theta^{0,0}.
\end{equation}
A power of $i$ has been inserted to make $\theta^{p,q}$ real:  The product of $n$ real fermions gets a sign $(-1)^{n(n-1)/2}$ under Hermitian conjugation, because of the reverse ordering.  The ghost $a$'s and $a^\dagger$'s are fermions, because they take fermions to bosons, and vice versa (in contrast to ordinary $\gamma$ matrices, which take fermions to fermions).  Thus $\theta^{p,q}$ is the product of $p+q+1$ fermions, including $\langle\theta|$ itself.
Then $\pi_{p,q}$ is not necessarily Hermitian, but has been defined to give 0 or 1 in graded commutators.  (But $\pi^{p,q}$ is always Hermitian, like $|\pi\rangle$ and $\pi_0$.)
We will sometimes also use a notation where $\theta^{p,q}$ carries instead $p$ $\oplus$'s and $q$ $\ominus$'s:  For example, $\theta^{1,0}\equiv\theta^{\oplus}$.
Note that the ghosts alternate in both statistics and chirality with each ghost level.

So in superspace notation the free super BRST operator is
\begin{equation}
\label{classq}Q'_{free}~=~\f{1}{2}c\,\square~-\f{1}{2}\bar{\pi}\gamma^{\oplus 2}\theta
b~+~\f{1}{4}\bar{\pi}\tilde{\gamma}^{\oplus}\pi~-\f{i}{2}\bar{\pi}\gamma^{\oplus}\rlap/p \theta,~~~~~\rlap/p~\equiv~-i\partial_{a}\gamma^{a}\\
\end{equation}
We can make a unitary transformation on $Q'_{free}$ to give a convenient
form with which to work. Specifically, the unitary transformation
\begin{equation}
\label{uniclassq} Q_{free}~=~U_0 Q'_{free}U_0^{\dag}\\
\end{equation}
with
\begin{equation}
\label{unitary}U_0~=~e^{\bar{\theta}\tilde{\gamma}^{\oplus}\theta b/2}\\
\end{equation}
gives $Q_{free}$ in terms of the supersymmetry generator $q_{0}$,
spinor covariant derivative $d_{0}$ and all their nonminimal versions:
\begin{eqnarray} \label{qad}
{Q}_{free}&=&\f{1}{2}c\,\square-2\bar{\pi}a^{\dagger\oplus}a^{\oplus}\theta b-\f{i}{2}\bar{q}a^{\dag\oplus}d\\
q&=&\pi-\rlap/p\theta ,\quad
d~=~\pi+\rlap/p \theta
\end{eqnarray}
Actually, $q_0$ is the only part of $q$ that does not appear in
this form of the BRST operator:  Because of the creation and
annihilation operators, $\theta_0$ and $\pi_0$ appear only as
their supersymmetry invariant combination $d_0$.  Thus the
supersymmetry generator that anticommutes with this form of $Q$ is
just the usual one $q_0$.  (This can also be derived in a
straightforward way by starting with the lightcone $q$.) Then the
supersymmetry generator for $Q'_{free}$ can be obtained by
inverting the unitary transformation on $q_{0}$:
\begin{eqnarray}
\label{qoriginal}
q'_{0}~&=&~U_0^{\dagger}q_{0}U_0 \nonumber\\
&=&\pi_{0}-\rlap/p\theta_{0}-\theta^{\oplus}b
\end{eqnarray}

\mysection{Interacting superparticle}

The (D=3,4,6,10) superparticle BRST operator in a super Yang-Mills
background (with constant superfield strength) is closely related
to the superstring BRST operator. The introduction of the SYM
background can be established by gauge covariantizing the super
covariant derivatives $p_{a}$ and $d_{0\alpha}$:
\begin{eqnarray}
\label{gauaging}p_{a}&\longrightarrow &\nabla_{a}\\
d_{0}&\longrightarrow & \nabla_{0}
\end{eqnarray}
Then the graded algebra among the covariant derivatives is
\cite{algebra}
\begin{eqnarray}
\label{algebra}
&&[\nabla_{a},\nabla_{b}]~=~F_{ab} \\
&&\{\nabla_{0 \alpha},\nabla_{0 \beta}\}~=~2\gamma_{a \alpha
\beta} \nabla^{a}\\
&&[\nabla_{0 \alpha},\nabla_{a}]~=~\gamma_{a \alpha
\beta}W^{\beta}
\end{eqnarray}

The Bianchi identity from the above algebra gives
\begin{equation}
\label{bianchi}\gamma_{a\alpha\beta}[\nabla^{a},W^{\beta}]~=~0
\end{equation}
and the D=3,4,6,10 dimensional gamma matrix~(which is symmetric in those cases)
identity
\begin{equation}
\gamma_{a(\alpha\beta}{{\gamma^{a}}_{\gamma)}}^{\delta}~=~0.
\end{equation}
We begin at linear order in the fields, where
the background satisfies the equations of motion
\begin{eqnarray}
\label{eom} &&\{\nabla_{\alpha},W^{\alpha}\}~=~0\\
&&[\nabla^{a},F_{ab}]~=~0
\end{eqnarray}

\mysubsection{Constant YM background}

One way to build this interacting super BRST operator is
by considering an ordinary constant YM background first, and next supersymmetrizing it by
including a constant fermionic field strength~(not yet superfield) $\on\circ w{}^{\alpha}$.
Then we extend the result to a nonconstant SYM background in the next subsection.

Making the gauge choice
$$ \on\circ A{}_a=\f{i}{2}x^{b}\on\circ F{}_{ba} $$
for constant field strength,
the super BRST operator can be written in the form
\begin{equation}
\label{ordiym}
\on\circ Q{}'_{YMB}~=~Q'_{free}~+~ \f12\on\circ F{}_{ab}V^{ab}\\
\end{equation}
We then find
\begin{equation}
V^{ab}\ =\
\f{i}2 cx^{[a}p^{b]}
~+~\f{i}2 R^{\oplus}R^{[a}p^{b]}
-~(c+R^{\oplus})~\bar{\pi}\gamma^{ab}\theta
~+~\f14 (x+R)^{[a}\bar{\pi}\gamma^{\oplus}\gamma^{b]}\theta
\end{equation}
in terms of an expression $R^i$ defined below,
where we use the notation
\begin{eqnarray}
C^{[a}D^{b]} &\equiv&~C^{a}D^{b}-C^{b}D^{a}\\
\gamma^{ab}&\equiv&~-\f{1}{4}\gamma^{[a}\gamma^{b]}
\end{eqnarray}
We can also write
\begin{eqnarray}
& V^{ab} = V^{\oplus ab} &\nonumber\\
& V^{ijk}=
i(x^i x^j+R^i R^j)p^{k} +
\f12(x+R)^{(i}\bar{\pi}\gamma^{j]}\gamma^{k}\theta
&\nonumber\\
& x^{\oplus}=c,~p^{\oplus}=0 &
\end{eqnarray}
(There is further antisymmetrization in the last two indices upon contraction with $F$, following the graded symmetrization in the first two indices shown above:  The tensor $V^{ijk}$ has mixed symmetry.)

The expression $R^i$ is given by
\begin{equation}
\label{defRG}
R^i(\theta)\equiv\f12\bar{\theta}\mathcal{O}\gamma^i\theta
\end{equation}
where the operator $\mathcal{O}$ is defined to satisfy
\begin{eqnarray}\label{condition}
~[\gamma^{\oplus},\mathcal{O}]&=&0\nonumber\\
~\{\tilde{\gamma}^{\oplus},\mathcal{O}\}&=&2\gamma^{\oplus}\nonumber\\
~[a^{\dag\oplus}a_{\oplus}-a^{\dag\ominus}a_{\ominus},\mathcal{O}]&=&0\nonumber\\
\langle0|\mathcal{O}\tilde{\gamma}^{\oplus}&=&-i\langle0|\tilde{\gamma}^{\oplus}~~=~~\langle0|\gamma^{\oplus}
\end{eqnarray}
As an explicit form of $\mathcal{O}$ we find
\begin{equation}
\mathcal{O}=\frac{1}{2}\left\{\frac{1}{\tilde{\gamma}^{\oplus}},\gamma^{\oplus}\right\}
\end{equation}
where
\begin{eqnarray}
\frac{1}{\tilde{\gamma}^{\oplus}}&=&\sum^{\infty}_{p=0}\left[\Theta_{N_{\oplus}-N_{\ominus}}\frac{N_{\oplus}!}{(N_{\oplus}+p+1)!}~ia_{\oplus}(ia_{\oplus}a_{\ominus})^{p}\right.\nonumber\\
&&\left.~~~~~~~~~~~~~~~~~~~~~~~-a^{\dag\ominus}(-ia^{\dag\oplus}a^{\dag\ominus})^{p}\Theta_{N_{\ominus}-N_{\oplus}}\frac{N_{\ominus}!}{(N_{\ominus}+p+1)!}\right]
\end{eqnarray}
with
$$ \Theta_{x}=\left\{\begin{matrix} ~1&~x\geq0 \\
~0&~x<0\end{matrix}\right. \quad, \quad\quad
N_\mu=a^{\dag\mu}a_{\mu} $$ (not summed over $\mu$). This
representation satisfies (\ref{condition}) if we regularize
indefinite norm states. (See the Appendices for details.)

\mysubsection{Constant SYM background}

 From this $\on\circ Q{}'_{YMB}$ we can construct a BRST operator
for a supersymmetric constant SYM background
$\on\circ Q{}'_{SYMB}$ in the form
\begin{equation}
\on\circ Q{}'_{SYMB}~=~Q'_{free}+\f12\on\circ F{}_{ab}V^{ab}+\on\circ w{}^\alpha V_{\alpha}
\end{equation}
In addition to first-quantized transformations we take
$q'_{0}$ (\ref{qoriginal}) to also generate the
second-quantized transformations of $\on\circ w{}^\alpha$ and
$\on\circ F{}_{ab}$
\begin{eqnarray}
\{q'_{0\beta},\on\circ w{}^\alpha\}&=&~{{\gamma^{ab}}_{\beta}}^{\alpha}\on\circ F{}_{ab}\\
\{q'_{0},\on\circ F{}_{ab}\}&=&0
\end{eqnarray}
so that they cancel up to a gauge transformation (generated by $Q'_{free}$):
\begin{eqnarray}
\{q'_{0},~\on\circ Q{}'_{SYMB}\}~&=&~\{Q'_{free},\Psi\}
\end{eqnarray}
We then have
\begin{equation}
\f12\on\circ F{}_{ab}~\{q'_{0\beta},V^{ab}\}~-~\on\circ w{}^\alpha[q'_{0\beta},V_{\alpha}]
~=~-\on\circ F{}_{ab}~{{\gamma^{ab}}_{\beta}}^{\alpha}~V_{\alpha}~+~\{Q'_{free},\Psi_{\beta}\}
\end{equation}
This is true if we define $V_{\alpha}$ by
\begin{equation}
\{q'_{0\beta},V^{ab}\}|_{\gamma^{ab}}~\equiv~-2\gamma^{ab}{}_{\beta}{}^{\alpha}~V_{\alpha}
\end{equation}
which means we define $V_{\alpha}$ from the left-hand side by
selecting only terms with an explicit $\gamma^{ab}$.

With this definition we find $V_{\alpha}$ and $\Psi_{\alpha}$
\begin{eqnarray}
V_{\alpha}~&\equiv&~-(c+R^{\oplus})~q'_{0\alpha}\\
\Psi_{\alpha}~&\equiv&~-\f{i}{2}(x^{b}+~R^{b})
(\gamma^{a}_{\alpha\beta}\theta_{0}^{\beta}\on\circ
F{}_{ab}-2\gamma_{b\alpha\beta}\on\circ w{}^{\beta})
+~\f{1}{2}(\gamma^{b}_{\alpha\beta}\theta_{0}^{\beta}+\gamma^{b}_{\alpha\beta}\tilde{\theta}_{0}^{\beta})
iR^{a}\on\circ F{}_{ab}\qquad\quad
\end{eqnarray}
where
\begin{equation}
\tilde{\theta}~=~-i\langle0|\mathcal{O}|
\theta\rangle~=~\langle0|e^{ia_{\oplus}a_{\ominus}}-1|
\theta\rangle
\end{equation}
which contains all nonminimal ghost-number-zero ghosts.

To obtain a gauge independent and explicitly supersymmetric expression
we perform a unitary transformation
$$
U_1~=~e^{\Lambda}
$$
where
\begin{eqnarray}
\Lambda~&=&iR^{b}~(~\theta_{0}\gamma_{b}\on\circ w
+~\tilde{\theta}\gamma_{b}\on\circ w
+~\f{1}{2}\theta_{0}\gamma_{b}\gamma^{ac}\theta_{0}\on\circ
F{}_{ac}
+~\f{1}{2}\tilde{\theta}\gamma_{b}\gamma^{ac}\theta_{0}\on\circ F{}_{ac}\nonumber\\
&&~~~~~~~~+~\f{1}{2}\tilde{\theta}\gamma_{b}\gamma^{ac}\tilde{\theta}\on\circ F{}_{ac}~-~\f{1}{2}\theta_{0}\gamma^{c}\tilde{\theta}\on\circ F{}_{bc})\nonumber\\
&&-~\theta_{0}\gamma^{b}\tilde{\theta}~(~\f{1}{3}\theta_{0}\gamma_{b}\on\circ w
+~\f{2}{3}~\tilde{\theta}\gamma_{b}\on\circ w
+~\f{1}{4}\theta_{0}\gamma_{b}\gamma^{ac}\theta_{0}\on\circ F_{ac}\nonumber\\
&&~~~~~~~~~+~\f{3}{4}~\tilde{\theta}\gamma_{b}\gamma^{ac}\theta_{0}\on\circ F_{ac}
+~\f{5}{12}~\tilde{\theta}\gamma_{b}\gamma^{ac}\tilde{\theta}\on\circ
F_{ac})
\end{eqnarray}
After another unitary transformation $U_0$
(\ref{unitary}),
$\on\circ Q{}'_{SYMB}$ becomes (at the linearized level)
\begin{eqnarray}
\label{sym}
\on\circ Q{}_{SYMB}&=&\f{1}{2}~(c+R^{\oplus}+\f{1}{2}\bar{\theta}\tilde{\gamma}^{\oplus}\theta~)|_{>}(~\Box~+~W\nabla_{0}~-~\bar{\pi}\gamma^{ab}\theta|_{>}~F_{ab}~)\hspace{3cm}\nonumber\\
&-&2~\bar{\pi}a^{\dagger\oplus}a^{\oplus}\theta|_{>}b~+~\f{1}{4}~\bar{\pi}\tilde{\gamma}^{\oplus}\pi|_{>}\nonumber\\
&-&\f{i}{2}~(\nabla_{a}~+\tilde{\theta}\gamma_{a}W~+\f{1}{2}~\tilde{\theta}\gamma_{a}\{W,\nabla_{0}\}\tilde{\theta}~)~\bar{\pi}\gamma^{\oplus}\gamma^{a}\theta|_{>}\nonumber\\
&-&\f{1}{2}~\bar{\pi}^{\oplus}\nabla_{0}~-~\f{1}{2}~(\nabla_{a}~+\tilde{\theta}\gamma_{a}W~+\f{1}{2}~\tilde{\theta}\gamma_{a}\{W,\nabla_{0}\}\tilde{\theta}~-~\f{i}{2}R^{b}|_{>}F_{ab})~\bar{\theta}^{\oplus}\gamma^{a}\nabla_{0}\nonumber\\
&-&\f{1}{3}~\bar{q}^{\oplus}\gamma^{a}\tilde{\theta}~\tilde{\theta}\gamma_{a}W~+~\f{5}{24}~\bar{q}^{\oplus}\gamma^{a}\tilde{\theta}~\tilde{\theta}\gamma_{a}\{W,\nabla_{0}\}\tilde{\theta}\nonumber\\
&+&\f{i}{4}~\bar{q}^{\oplus}\gamma^{b}\tilde{\theta}~R^{a}|_{>}~F_{ab}\nonumber\\
&+&\f{1}{4}~[~i\nabla_{b}R^{b}|_{>},-i\nabla_{a}~\bar{\pi}\gamma^{\oplus}\gamma^{a}\theta|_{>}]\nonumber\\
&+&~R^{\oplus}|_{>}P_{a}(\tilde{\theta}\gamma^{a}W~+~\f{1}{2}\tilde{\theta}\gamma^{a}\{W,\nabla_{0}\}\tilde{\theta})\nonumber\\
&+&\f{1}{2\cdot3!}~\left[~i\nabla_{c}R^{c}|_{>},[~i\nabla_{b}R^{b}|_{>},-i\nabla_{a}~\bar{\pi}\gamma^{\oplus}\gamma^{a}\theta|_{>}]\right]
\end{eqnarray}
where~ $|_{>}$ means that we drop $\theta_{0}$ contributions, and

\begin{eqnarray}
\Box&=&-\nabla^{a}\nabla_{a}\\
\nabla_{a}&=&p_{a}~+~A_{a}\\
\nabla_{0\alpha}&=&\pi_{0\alpha}~+~(\rlap/p\theta_{0})_{\alpha}~+~A_{\alpha}\\
q^{\oplus\alpha}&=&\pi^{\oplus\alpha}~+~(\rlap/p\theta^{\oplus})^{\alpha}
\end{eqnarray}
The superfields have the $\theta_0$ expansions
\begin{eqnarray}
F_{ab}~&=&~\on\circ F{}_{ab}\\
W^{\alpha}&=&\on\circ w{}^{\alpha}+~(\gamma^{ab}\theta_{0})^{\alpha}~\on\circ F{}_{ab}\\
A_{a}&=&\on\circ A{}_a~+~\bar{\theta}_{0}\gamma_{a}\on\circ w+~\f{1}{2}~\bar{\theta}_{0}\gamma_{a}\gamma^{bc}\theta_{0}~\on\circ F{}_{bc}\\
A_{\alpha}&=&~(\gamma^{a}\theta_{0})_{\alpha}\on\circ A{}_a~+~\f{2}{3}(\gamma^{a}\theta_{0})_{\alpha}~\bar{\theta}_{0}\gamma_{a}\on\circ w+~\f{1}{4}~(\gamma^{a}\theta_{0})_{\alpha}~\bar{\theta}_{0}\gamma_{a}\gamma^{bc}\theta_{0}~\on\circ F{}_{bc}\qquad
\end{eqnarray}
in the gauge
\begin{eqnarray}
\on\circ A{}_a~&=&\f{i}{2}x^{b}~\on\circ F{}_{ba}\\
\on\circ A{}_\alpha ~&=&~0
\end{eqnarray}
used above,
but (\ref{sym}) is manifestly gauge independent and
supersymmetric.

\mysubsection{Arbitrary SYM background}

%\begin{eqnarray}
%\label{arbsym} Q_{SYMB}&=&~e^{\tilde{\theta}\nabla_{0}}~e^{R^{a}|_{>}\nabla_{a}}\times~\nonumber\\
%&&\left[~\f{1}{2}~\left(c+R^{\oplus}+\f{1}{2}\bar{\theta}\tilde{\gamma}^{\oplus}\theta~\right)|_{>}\left(~\square+W\nabla_{0}-\bar{\pi}\gamma^{ab}\theta|_{higher}~F_{ab}~\right)\right.\nonumber\\
%&&\left.-2~\bar{\pi}a^{\dagger\oplus}a^{\oplus}\theta~b~+~\f{1}{4}~\bar{\pi}\tilde{\gamma}^{\oplus}\pi|_{>}~\right]~e^{-\f{i}{2}R^{a}|_{>}\nabla_{a}}~e^{-\tilde{\theta}\nabla_{0}}
%\end{eqnarray}
After making a final unitary transformation
$$
U_3 =
e^{(R^{\oplus}+\theta\tilde{\gamma}^{\oplus}\theta/2)|_{>}b}
$$
the above BRST operator can be written in the simple form
$$
Q''_{SYMB} =~U
\left[~\f{1}{2}c\left(~\square+W\nabla_{0}-\bar{\pi}\gamma^{ab}\theta|_{>}~F_{ab}~\right)\right.
+\left.\f{1}{4}~\bar{\pi}\tilde{\gamma}^{\oplus}\pi|_{>}~\right]
U^{-1}
$$
\begin{equation}
\label{arbsymsimple} U =
e^{\tilde{\theta}\nabla_{0}}~e^{iR^{a}|_{>}\nabla_{a}}~e^{(R^{\oplus}+\theta\tilde{\gamma}^{\oplus}\theta/2)|_{>}b}
\end{equation}
which can be applied directly to the case of an arbitrary, nonlinear SYM
background.

In fact, the nilpotence of this BRST operator does not seem to require that the background be on shell.  This is contradictory to the usual result that any description of linearized ``quantum" Yang-Mills in a Yang-Mills background must have the background on shell, since nonabelian gauge invariance relates kinetic and interaction terms \cite{Deser}.  (Similar remarks apply to any nonabelian gauge theory, such as gravity or strings.)  This paradox is probably due to the fact that we have not required an ``integrability" condition on the background:  For a generic self-interacting field theory, an action (or ZJBV action) of the form
\begin{equation}
S = \f12\phi^j\phi^i K_{ij} +\f16\phi^k\phi^j\phi^i V_{ijk} + ...
\end{equation}
results in the kinetic operator (or BRST operator) in a background
\begin{equation}
Q_{ij} = K_{ij} +\phi^k V_{kij} + ...
\end{equation}
 From $S$ we can see that $K,V,...$ must be totally (graded) symmetric.  In $Q$, this condition on $K$ is seen to follow simply from hermiticity, but the condition on $V$ is not so obvious.  Since we are ultimately concerned with the BRST operator for the superstring without background,  and are using the SYM case in a background only as an analogy, we will not consider this obscurity further here.

As explained in the Introduction, in the above expression for the BRST operator (\ref{arbsymsimple}) we are not allowed to remove the exponential factors, since that would lead to
a trivial result. This fact can be understood already in the free
case:  The BRST operator that would result from dropping the background and exponentials has the wrong cohomology, since the remaining two terms have no dependence on $\theta_0$, so one would obtain an ordinary superfield satisfying only the Klein-Gordon equation.  In this case the exponentials are required for $Q$ to be regularizable:  Certain poorly defined quantities cancel upon their expansion.  (See Appendices B-C.)

\mysection{Superstring}

The superstring
is described by a 2D field theory whose algebra of covariant derivatives (currents) resembles
that of interacting particle covariant derivatives for a constant SYM background:
\begin{eqnarray}
\{D^{(\pm)}_{\alpha}(1),D^{(\pm)}_{\beta}(2)\}&=&2\delta(2-1)\gamma^{a}_{\alpha\beta}P^{(\pm)}_{a}(1)\nonumber\\
~[D^{(\pm)}_{\alpha}(1),P^{(\pm)}_{a}(2)]&=&2\delta(2-1)\gamma_{a\alpha\beta}\Omega^{(\pm)\beta}(1)\nonumber\\
\{D^{(\pm)}_{\alpha}(1),\Omega^{(\pm)\beta}(2)\}&=&\pm i\delta'({2-1})\delta_{\alpha}^{\beta}\nonumber\\
~[P^{(\pm)}_{a}(1),P^{(\pm)}_{b}(2)]&=&\pm i\delta'(2-1)\eta_{ab}\nonumber\\
~[P^{(\pm)},\Omega^{(\pm)}] &=&\{\Omega^{(\pm)},\Omega^{(\pm)}\} \ = \ 0
\end{eqnarray}
where
\begin{eqnarray}
D^{(\pm)}_{\alpha}&=&\pi_{0\alpha}+(\gamma^{a}\theta_{0})_{\alpha}\hat{P}^{(\pm)}_{a}\pm i\f{1}{2}(\gamma^{a}\theta_{0})_{\alpha}\theta_{0}\gamma_{a}\theta'_{0}\nonumber\\
P^{(\pm)}_{a}&=&\hat{P}^{(\pm)}_{a}\pm i\theta_{0}\gamma_{a}\theta'_{0}\nonumber\\
\Omega^{(\pm)\alpha}&=&\pm i\theta'_{0}
\end{eqnarray}
and
\begin{equation}
\hat{P}^{(\pm)}=\f{1}{\sqrt{2}}\left(i\frac{\delta}{\delta X}\pm
X'\right)
\end{equation}
in the Hamltonian formalism correspond to the left(right)-moving combinations of $P_{0}$ and $P_{1}$ of the first-order formalism
after using the equation of motion for $P_1$ (see below).
(In the definitions above, $(\pm)$'s
on $\pi$ and $\theta$ are understood.)
Also, $'$ means a $\sigma$ derivative as usual.
$D,P,\Omega$ (anti)commute with the supersymmetry generator
\begin{equation}
q_\alpha\ =\ q^{(+)}_\alpha +q^{(-)}_\alpha ,\quad
q^{(\pm)}_{\alpha}\ =\ \int\pi_{0\alpha}-(\gamma^{a}\theta_{0})_{\alpha}\hat{P}^{(\pm)}_{a}\mp i\f{1}{6}(\gamma^{a}\theta_{0})_{\alpha}\theta_{0}\gamma_{a}\theta'_{0}
\end{equation}

So we can see the analogy between the covariant derivatives of the free
superstring and the superparticle with SYM
background.
\begin{equation}
(D_{\alpha},P_{a},\Omega^{\alpha})\leftrightarrow(\nabla_{\alpha},\nabla_{a},W^{\alpha})
\end{equation}
as well as the less precise analogy
\begin{equation}
{}' \leftrightarrow F_{ab}
\end{equation}

\mysubsection{BRST}

Now we can guess the result for the superstring BRST operator from the result of the
superparticle in a constant SYM background:
\begin{equation}
\label{sstringstart} Q_{sstring}=U\left(~\int
c~T~+~\f{1}{4}\bar{\pi}\tilde{\gamma}^{\oplus}\pi|_{>}~\right)U^{-1}
\end{equation}
where
\begin{equation}
U=e^{\int\tilde{\theta}D}~e^{\int
iR^{a}|_{>}P_{a}}~e^{\int(R^{\oplus}+\theta\tilde{\gamma}^{\oplus}\theta/2)|_{>}b}
\end{equation}
and  $Q_{sstring}=Q^{(+)}_{sstring}+Q^{(-)}_{sstring}$.
 From now on we will suppress the $\sigma$-integral symbol  for
convenience.  Since $\theta_0$ and $\pi_0$ appear only in $D$, $P$, and $T$, this $Q$ is automatically supersymmetric under the above supersymmetry generator.

There are two major differences in $T$ as compared to the superparticle:  Firstly the string has a $c'b$ ghost contribution. Secondly the superstring has $\Omega D$ as an analog of
$W\nabla_{0}-\bar{\pi}\gamma^{ab}\theta F_{ab}$, from the correspondence above.
So our trial form of $T$ is
\begin{eqnarray}
T^{(\pm)}&=&\f{1}{2}~\square_{\pm}\mp i \left\{c'b + \bar{\theta}'\pi + w^{\pm}(\bar{\theta}\pi)' \right.\nonumber\\
&&\left.+A^{\pm}_{1}\left[\bar{\theta}(a^{\dag\oplus}a_{\oplus}-a^{\dag\ominus}a_{\ominus})\pi\right]'
+A^{\pm}_{2}\left[\bar{\theta}(a^{\dag\oplus}a_{\oplus}+a^{\dag\ominus}a_{\ominus})\pi\right]'\ \right\}
\end{eqnarray}
where $\square_{\pm}=-\hat{P}^{(\pm) a}\hat{P}^{(\pm)}_{a}$.
(The true energy-momentum tensor is actually $T\mp i(cb)'$.)
The constants $w^{\pm}$,~$A^{\pm}_{1}$ and $A^{\pm}_{2}$ will be
determined by 3 conditions: (1) The conformal weight of
$\bar{\pi}\tilde{\gamma}^{\oplus}\pi$ should be $1$. (2) The
conformal anomaly should cancel in $D=10$. (3) $\theta_{0}$ should
have conformal weight $0$ due to supersymmetry.  (The $A_1$ term
is ghost number, while the $A_2$ term is ghost level.)

Satisfying these constraints we find
\begin{eqnarray}
A^{\pm}_{1} &=& 1 , \quad A^{\pm}_{2}\ =\ w^{\pm}\ =\ 0\nonumber\\
\Rightarrow\quad T^{(\pm)}&=&\f{1}{2}~\square_{\pm}\mp i \left\{\
c'b +\bar{\theta}'\pi
+\left[\bar{\theta}(a^{\dag\oplus}a_{\oplus}-a^{\dag\ominus}a_{\ominus})\pi\right]'\
\right\}
\end{eqnarray}
and the gauge-fixed Hamiltonian is $\{Q,\int b^{(+)}+b^{(-)}\}=\int T^{(+)}+T^{(-)}$.

This $Q$ has four interesting quantum numbers:(1) ghost number;
(2) conformal weight, which is ``momentum number" (1 for $P,b,\pi,{}'$) minus ghost
number; (3) (10D) engineering dimension
($-1$ for $x,c$, $-\f12$ for $\theta$, 2 for ${}'$); and (4) a
mysterious ``field weight", which is 1 for all fields, but for
which we attribute a 1 for $\tilde\gamma^\oplus$.  (Thus, $Q$ is
quadratic in momenta and primes, and cubic in fields and
$\tilde\gamma^\oplus$'s.)

 From now on let's concentrate on one chirality. After expanding the exponential factor and regularizing (as explained in Appendix C) we find
\begin{eqnarray}
\label{sstring}
Q^{(+)}_{sstring}&=&\left(c+R^{\oplus}+\f{1}{2}\bar{\theta}\tilde{\gamma}^{\oplus}\theta~\right)|_{>}\nonumber\\
&&~~~~~~~~~\times\left(~\f{1}{2}\square_{+}~-ic'b~-~i\bar{\theta}'\pi~-~i[\bar{\theta}(a^{\dag\oplus}a_{\oplus}-a^{\dag\ominus}a_{\ominus})\pi]'\right)\nonumber\\
&-&2~\bar{\pi}a^{\dagger
\oplus}a^{\oplus}\theta|_{>}b~+~\f{1}{4}~\bar{\pi}\tilde{\gamma}^{\oplus}\pi|_{>}\nonumber\\
&-&\f{i}{2}~\left(\hat{P}_{a}~+i\theta_{0}\gamma_{a}\theta'_{0}~+2i\tilde{\theta}\gamma_{a}\theta'_{0}~+~i\tilde{\theta}\gamma_{a}\tilde{\theta}'~-\f{1}{2}~R'_{a}|_{>}\right)~\bar{\pi}\gamma^{\oplus}\gamma^{a}\theta|_{>}\nonumber\\
&-&\f{1}{2}~\bar{\pi}^{\oplus}\left(\pi_{0}+(\gamma^{a}\theta_{0})\hat{P}_{a}+i\f{1}{2}(\gamma^{a}\theta_{0})\theta_{0}\gamma_{a}\theta'_{0}~\right)\nonumber\\
&-&\f{1}{2}~\left(\hat{P}_{a}~+i\theta_{0}\gamma_{a}\theta'_{0}~+2i\tilde{\theta}\gamma_{a}\theta'_{0}~+~i\tilde{\theta}\gamma_{a}\tilde{\theta}'~-\f{1}{2}~R'_{a}|_{>}\right)\nonumber\\
&&~~~~\times\bar{\theta}^{\oplus}\gamma^{a}\left(\pi_{0}+(\gamma^{b}\theta_{0})\hat{P}_{b}+i\f{1}{2}(\gamma^{b}\theta_{0})\theta_{0}\gamma_{b}\theta'_{0}~\right)\nonumber\\
&-&\f{1}{3}~\bar{q}^{\oplus}\gamma^{a}\tilde{\theta}\left(2i\tilde{\theta}\gamma_{a}\theta'_{0}~+~i\f{5}{4}~\tilde{\theta}\gamma_{a}\tilde{\theta}'-\f{3}{4}~R'_{a}|_{>}\right)\nonumber\\
&+&\f{1}{2}~\left(\hat{P}_{a}~+i\theta_{0}\gamma_{a}\theta'_{0}~+2i\tilde{\theta}\gamma_{a}\theta'_{0}~+~i\tilde{\theta}\gamma_{a}\tilde{\theta}'~-~\f{1}{2}R'_{a}|_{>}\right)^{2}R^{\oplus}|_{>}\hspace{0.7cm}\nonumber\\
&+&ic'b~\left(R^{\oplus}+\f{1}{2}\bar{\theta}\tilde{\gamma}^{\oplus}\theta\right)|_{>}
\end{eqnarray}
This $Q$ satisfies $Q^2=0$, as can be checked directly.

\mysubsection{Constraints}

The constraints of the gauge-invariant action (see following
subsection) can be obtained directly from the BRST operator by
taking its (graded) commutator and keeping just ghost-number-zero
terms: The Virasoro constraints $\mathcal{A}$ follow as usual from
$b$ (with the gauge-fixed action from $\int b$), while
generalizations $\mathcal{B}$ of the $\gamma\cdot pd$ constraint
($\kappa$ symmetry generator) follow from $\theta^{p,p+1}$, and
first-class generalizations $\mathcal{E}$ of the second-class
constraint $d$ follow from $\pi^{p,p+1}$ \cite{ilk}:
\begin{eqnarray}
\mathcal{A}&=&~\f{1}{2}\square_{+}-i\sum_{q=0}^\infty\bar{\theta}'^{q,q}\pi_{q,q}\\
\mathcal{B}_{0}&=&\gamma^{a}\Pi\mathcal{P}_{a}(\textbf{\emph{d}}_{0}+\pi^{1,1})-2\theta^{1,1}\mathcal{A}+2\vartheta^{0}(\f{1}{2}\mathcal{P}^{2}+\mathcal{A})\\
\mathcal{B}_{p}&=&\gamma^{a}\Pi\mathcal{P}_{a}(\pi^{p,p}+\pi^{p+1,p+1})+~2(\theta^{p,p}-\theta^{p+1,p+1})\mathcal{A}
+2\vartheta^{p}(\f{1}{2}\mathcal{P}^{2}+\mathcal{A})\qquad\\
\mathcal{E}_{0}&=&\textbf{\emph{d}}_{0}-\pi^{1,1}+\gamma^{a}\Pi\mathcal{P}_{a}\theta^{1,1}\\
\mathcal{E}_{p}&=&~\Pi(\pi^{p,p}-\pi^{p+1,p+1})+\gamma^{a}\Pi\mathcal{P}_{a}(\theta^{p,p}+\theta^{p+1,p+1})
\end{eqnarray}
where
\begin{eqnarray}\label{vartheta}
\mathcal{P}_{a}&\equiv&\hat{P}_{a}~+i\theta_{0}\gamma_{a}\theta'_{0}~+2i\tilde{\theta}\gamma_{a}\theta'_{0}~+~i\tilde{\theta}\gamma_{a}\tilde{\theta}'~-\f{1}{2}~\tilde{R}'_{a}|_{>}\nonumber\\
\textbf{\emph{d}}_{0}&\equiv&\pi_{0}+\gamma^{a}\hat{P}_{a}\theta_{0}+i\f{1}{2}(\gamma^{a}\theta_{0})\theta_{0}\gamma_{a}\theta'_{0}~+\f{2}{3}\gamma^{a}\tilde{\theta}\left(2i\tilde{\theta}\gamma_{a}\theta'_{0}~+~i\f{5}{4}~\tilde{\theta}\gamma_{a}\tilde{\theta}'-\f{3}{4}~\tilde{R}'_{a}|_{>}\right)\nonumber\\
\vartheta^{p}&\equiv&2\theta^{1,1}+2\theta^{2,2}+\cdots+2\theta^{p-1,p-1}+\theta^{p,p}-\theta^{p+1,p+1}-2\theta^{p+2,p+2}-2\theta^{p+3,p+3}-\cdots\nonumber\\
\end{eqnarray}
and $\tilde{R}^{a}$ indicates that only ghost-number-zero $\theta$'s
are selected.

Since this procedure requires the component expression,
we explicitly use the projection operator
\begin{equation}
\Pi=\frac{1}{\tilde{\gamma}^{\oplus}}(\tilde{\gamma}^{\oplus})_{reg}
\end{equation}
as, e.g., $\Pi|\theta\rangle$, in some terms, as explained in
Appendix C. Also, because $R^{\oplus}$ only interacts with
$\pi\Pi$ we express $[\pi^{p,p+1},R^{\oplus}]$ as a projected
expression $\vartheta^{p}$. (For the full expression, see Appendix
D.)

These constraints are closed classically (after regularization: see Appendix D)
\begin{eqnarray}
~[\mathcal{A}(1),\mathcal{A}(2)]&=&-\delta'(2-1)[\mathcal{A}(1)+\mathcal{A}(2)]\nonumber\\
~[\mathcal{A}(1),\mathcal{E}_{p}(2)]&=&-\delta'(2-1)\mathcal{E}_{p}(1)\nonumber\\
~[\mathcal{A}(1),\mathcal{B}_{p}(2)]&=&-\delta'(2-1)[\mathcal{B}_{p}(1)+\mathcal{B}_{p}(2)]\nonumber\\
~\{\mathcal{E}_{p}(1),\mathcal{E}_{q}(2)\}&=&0\nonumber\\
~\{\mathcal{B}^{\alpha}_{p}(1),\mathcal{B}^{\beta}_{q}(2)\}
&=&8\delta'(2-1)\vartheta^{p\alpha}\mathcal{P}_{a}(1)\gamma^{a\beta\delta}E_{q\delta}(2)\nonumber\\
&&+4\delta'(2-1)\gamma^{\alpha\lambda}_{a}(\pi^{p,p}+\pi^{p+1,p+1})_{\lambda}(1)\gamma^{a\beta\delta}E_{q\delta}(2)\nonumber\\
&&-2\delta'(2-1)(\theta^{p,p}-\theta^{p+1,p+1}+\vartheta^{p})^{\alpha}\mathcal{B}^{\beta}_{q}[(1)+(2)]\nonumber\\
&&+((p,\alpha,1)\leftrightarrow (q,\beta,2))\nonumber\\
\{\mathcal{E}_{p\alpha}(1),\mathcal{B}^{\delta}_{q}(2)\}&=&-4\delta^{\beta}_{\alpha}\delta(2-1)\mathcal{A}(1)\nonumber\\
&&-2\delta'(2-1)(\theta^{q,q}-\theta^{q+1,q+1}+\vartheta^{p})^{\beta}\mathcal{E}_{p\alpha}(2)
\end{eqnarray}
where
\begin{eqnarray*}
E_{q}&\equiv&\sum_{r}i^{r-q}\frac{1}{\sqrt{r}}(\Theta_{r-q-1}-\Theta_{r-q-2})\mathcal{E}_{r-1}
\end{eqnarray*}

\mysubsection{Action}

 From this Hamiltonian form of the
BRST operator for the superstring we can find
the ZJBV form
\cite{ZJBV}:
\begin{equation}
Q^{ZJBV}=\f{i}{2}\dot{\phi}^{A}\phi_{A}-H-\check{\phi}_{A}\{\phi^{A},Q^{H}]
\end{equation}
where
\begin{eqnarray}
[\phi^{A},\phi^{B}\}&=&\Omega^{AB}\nonumber\\
\phi^{A}=\Omega^{AB}\phi_{B},&&\phi_{B}=\phi^{A}\Omega_{AB}
\end{eqnarray}
and $\check{\phi}_{A}$ is the antifield which is canonically
conjugate to the field $\phi^{A}$ by the antibracket.
ZJBV is useful for Lagrangian quantization, but since $Q$ is sufficient for Hamiltonian quantization, we leave the details for Appendix E.

Constraints appearing in the gauge-invariant Hamiltonian have ghost number 0; their ghosts have ghost number 1; their antighosts have ghost number $-1$; the antifields of their antighosts have ghost number $0$, and we can identify them in the ZJBV BRST operator with the Lagrange multipliers of the gauge-invariant Hamiltonian. (More generally, we interpret all the negative-ghost-number fields as antifields.) Let
\begin{eqnarray}
\Phi_{p,p+1}~&\equiv&~(-1)^{p+1}i\f{1}{2}\sqrt{p+1}\check{\theta}_{p,p+1}\nonumber\\
\Psi_{p,p+1}~&\equiv&~(-1)^{p+1}i\f{1}{2}\sqrt{p+1}\check{\pi}_{p,p+1}\nonumber\\
\check{b}~&\equiv&~g
\end{eqnarray}
and similarly for their antifields. Then we find the gauge
invariant action in Hamiltonian form $S_H$ either from the usual Hamiltonian procedure (using the constraints of the previous subsection), or as the antifield-free
part of $Q^{ZJBV}_{sstring}$:
\begin{equation}
S_H=-\dot{X}\hat
P^{0}+i\sum_{\pm,p}\dot{\theta}^{p,p}\pi_{p,p}-\sum_{\pm}~g_{\pm}\mathcal{A}_{\pm}+\sum_{\pm,p\geq0}\left(\tilde{\Phi}_{\pm,p,p+1}\mathcal{E}_{p}+~\tilde{\Psi}_{\pm,p,p+1}\mathcal{B}_{p}\right)
\end{equation}
(Again, for each sign $\pm$ we use fermions of the corresponding
chirality.)

If we consider only the quadratic terms and the $\mathcal{A}$ term, keeping only the physical fields, and
introducing $\hat P^1$ as an independent variable,
we find the first-order, 2$D$ world-sheet
covariant form \cite{affine} (with world-sheet metric $\eta_{mn}=(-+)$)
\begin{equation}
\label{2dcovariant}S^{phys}_{0}\ =\ \hat{P}^{m}\partial_{m}X-\f{1}{2}g_{mn}\hat{P}^{m}\hat{P}^{n}+i\sqrt{2}\sum_{\pm}\partial_{\pm}\theta^{\pm}_{0}\pi^{\pm}_{0}
\end{equation}
where $\theta^{\pm}_{0}\equiv\theta_{0L,R}$,
$\pi^{\pm}_{0}\equiv\pi_{0L,R}$, and
$\partial_{\pm}\equiv\f{1}{\sqrt{2}}(e_0{}^m\pm
e_1{}^m)\partial_m$. By introducing supersymmetric variables
\begin{eqnarray}
P^{m}&=&\hat{P}^{m}+\epsilon^{mn}(\eta^{+}_{(0)n}-\eta^{-}_{(0)n})\nonumber\\
D^{\pm}_{0}&=&\pi^{\pm}_{0}+[\hat{P}^{\pm}\pm\f{1}{2}(\eta^{+}_{(0)\mp}-\eta^{-}_{(0)\mp})]\cdot\gamma\theta^{\pm}_{0}\nonumber\\
\eta^{\pm}_{(0)m}&\equiv&\f{i}{\sqrt{2}}(\partial_{m}\theta^{\pm}_{0})\gamma\theta^{\pm}_{0}
\end{eqnarray}
(where we suppress spacetime indices for simplicity) and plugging
this into (\ref{2dcovariant}) we find
\begin{eqnarray}
S^{phys}_{0}&=&-\f{1}{2}g_{mn}P^{m}P^{n}+P^{m}[\partial_{m}X-(\eta^{+}_{(0)m}+\eta^{-}_{(0)m})]\nonumber\\
&&-\epsilon^{mn}[(\partial_{m}X)\cdot(\eta^{+}_{(0)n}+\eta^{-}_{(0)n})-\eta^{+}_{(0)m}\eta^{-}_{(0)n}]+i\sqrt{2}\sum_{\pm}\partial_{\pm}\theta^{\pm}_{0}D^{\pm}_{0}\qquad
\end{eqnarray}

Except for the last term this is the Green-Schwarz supersting
action. To extend this redefinition to the whole action one can
further define (we use $p,q$ for ghost level and $(p),(q)$ when
there is confusion with world sheet indices $l,m,n$)
\begin{eqnarray}
\mathcal{P}^{m}&=&\hat{P}^{m}+\epsilon^{mn}(\eta^{+}_{(0)n}-\eta^{-}_{(0)n})+\epsilon^{mn}(\chi^{+}_{n}-\chi^{-}_{n})\nonumber\\
\mathcal{D}^{\pm}_{0}&=&\pi^{\pm}_{0}+\left\{[\hat{P}^{\pm}\pm
\f{1}{2}(\eta^{+}_{(0)\mp}-\eta^{-}_{(0)\mp})]\cdot\gamma\theta^{\pm}_{0}
\mp\f{2}{3}(\xi^{+}_{\mp}-\xi^{-}_{\mp})\cdot\gamma\tilde{\theta}^{\pm}\right\}\nonumber\\
\mathcal{D}^{\pm}_{p}&=&\pi^{\pm}_{p}+\mathcal{P}^{\pm}\cdot\gamma\theta^{\pm}_{p}~~~~~~~~~(p\geq1)\nonumber\\
\pi^{\pm}_{p}&\equiv&\pi^{p,p}_{L,R}\nonumber\\
\theta^{\pm}_{p}&\equiv&\theta^{p,p}_{L,R}\nonumber\\
\vartheta^{\pm}_{p}&\equiv&\vartheta^{p}_{L,R}\nonumber\\
\chi^{\pm}_{m}&\equiv&\f{1}{\sqrt{2}}\left(2i(\partial_{m}\theta^{\pm}_{0})\gamma\tilde{\theta}^{\pm}-i(\partial_{m}\tilde{\theta}^{\pm})\gamma\tilde{\theta}^{\pm}+\f{1}{2}\partial_{m}\tilde{R}^{\pm}\right)\nonumber\\
\xi^{\pm}_{m}&\equiv&\f{1}{\sqrt{2}}\left(2i(\partial_{m}\theta^{\pm}_{0})\gamma\tilde{\theta}^{\pm}-i\f{5}{4}(\partial_{m}\tilde{\theta}^{\pm})\gamma\tilde{\theta}^{\pm}+\f{3}{4}\partial_{m}\tilde{R}^{\pm}\right)\nonumber~~~~~~~~~~~~~~~~~~\\
\eta^{\pm}_{(p)m}&\equiv&\f{i}{\sqrt{2}}(\partial_{m}\theta^{\pm}_{p})\gamma\theta^{\pm}_{p}\nonumber\\
\Phi_{p\pm}&\equiv&~\Phi^{L,R}_{p,p+1}\nonumber\\
\Psi_{p\pm}&\equiv&~\Psi^{L,R}_{p,p+1}
\end{eqnarray}

Then our manifestly worldsheet-covariant action reads
\begin{equation}
S_{0}=\tilde{S}_{GS}+i\sqrt{2}\sum_{\pm,p\geq0}\partial_{\pm}\theta^{\pm}_{p}\mathcal{D}^{\pm}_{p}+S_{A}
\end{equation}
where $S_{A}$ consists of Lagrange multipliers times all the (first-class) constraints other than Virasoro
\begin{equation}
S_{A}\ =\
\sum_{\pm,p\geq0}\left(\Psi_{p\pm}\mathcal{B}_{p}^\pm
+\Phi_{p\pm}\mathcal{E}_{p}^\pm\right)
\end{equation}
and $\tilde{S}_{GS}$ is an extension of the usual $GS$ action to
the fields $\theta^\pm_p$ at nonzero ghost levels
\begin{eqnarray}
\tilde{S}_{GS}&=&-\f{1}{2}g_{mn}\mathcal{P}^{m}\mathcal{P}^{n}\nonumber\\
&&+\sum_\pm\mathcal{P}^\pm\left[\partial_\pm X-(\eta^{+}_{(0)\pm}+\eta^{-}_{(0)\pm})+\sum_{p\geq1}\eta^{\pm}_{(p)\pm}
\pm(\chi^{+}_\pm-\chi^{-}_\pm)\right]\nonumber\\
&&-\epsilon^{mn}\left[(\partial_{m}X)\cdot\left\{(\eta^{+}_{(0)n}-\eta^{-}_{(0)n})+(\chi^{+}_{n}-\chi^{+}_{n})\right\}+\eta^{+}_{(0)m}\eta^{-}_{(0)n}\right]\nonumber\\
&&+\f{2}{3}\sum_{\pm}\pm\f{i}{\sqrt{2}}(\partial_{\pm}\theta^{\pm}_{0}\gamma\tilde{\theta}^{\pm})(\xi^{+}_\mp-\xi^{-}_\mp)
\end{eqnarray}
This is a first-order action in terms of the coordinates $X,
\theta_p^\pm$, momenta $\mathcal{P}^m, \mathcal{D}_p^\pm$,
worldsheet metric $g_{mn}$, and Lagrange multipliers
$\Phi_{p\pm},\Psi_{p,\pm}$.  Now $\mathcal{E}_{p}$ and
$\mathcal{B}_{p}$ are expressed in terms of these new variables as
\begin{eqnarray}
\mathcal{E}_{p}^\pm&=&\mathcal{D}^{\pm}_{p}-\mathcal{D}^{\pm}_{p+1}+2\mathcal{P}^{\pm}\cdot\gamma\theta^{\pm}_{p+1}\nonumber\\
\mathcal{B}_{p}^\pm&=&\mathcal{P}^{\pm}\cdot\gamma(\mathcal{D}^{\pm}_{p}+\mathcal{D}^{\pm}_{p+1}-2\mathcal{P}^{\pm}\cdot\gamma\theta^{\pm}_{p+1})\nonumber\\
&+&(\Theta_{p-1}\theta^{\pm}_{p}-\theta^{\pm}_{p+1}+\vartheta^{\pm}_{p})\times\left\{\frac{}{}\mathcal{P}^{\pm2}-[\mathcal{P}^{\pm}\mp(\eta^{+}_{(0)\mp}-\eta^{-}_{(0)\mp}+\chi^{+}_\mp-\chi^{-}_\mp)]^{2}\right.\nonumber\\
&&\left.\pm \f{1}{\sqrt{2}}\left[\sum_{q\geq1}i\partial_\pm\theta^{\pm}_{q}\left(\mathcal{D}^{\pm}_{q}-\mathcal{P}^{\pm}\cdot\gamma\theta^{\pm}_{q}\right)+i\partial_\pm\theta^{\pm}_{0}\right.\right.\nonumber\\
&&\left.\left.\times\left(\mathcal{D}^{\pm}_{0}-\left\{[\hat{P}^{\pm}\pm\f{1}{2}(\eta^{+}_{(0)\mp}-\eta^{-}_{(0)\mp})]\cdot\gamma\theta^{\pm}_{0}\mp\f{2}{3}(\xi^{+}_\mp-\xi^{-}_\mp)\cdot\gamma\tilde{\theta}^{\pm}\right\}\right)~\right]~~\right\}\nonumber\\
\end{eqnarray}
Elimination of $\mathcal{P}^1$ by its equation of motion reproduces the previous Hamiltonian form of the action except for terms quadratic in $\mathcal{E}$, which can be eliminated by a redefinition of $\Phi$.
The gauge-fixed action with ghosts is most easily obtained from the Hamiltonian formalism as $H=\{Q,\int b\}=\int T$:  Then (with the full $\theta^{p,q}$)
\begin{equation}
S_{GF}\ =\
\hat{P}^{m}\partial_{m}X-\f{1}{2}\eta_{mn}\hat{P}^{m}\hat{P}^{n}
+i\sqrt{2}\sum_{\pm}\partial_{\pm}c^{\pm}b_{\pm\pm}
+i\sqrt{2}\sum_{\pm}\partial_{\pm}\theta^{\pm}\pi^{\pm}
\end{equation}

\mysection{Future}

We have given a gauge-invariant action for the superstring and its corresponding BRST operator.  The BRST-invariant gauge-fixed action is the obvious quadratic expression following from $\{ Q , \int b \}$ (and is thus BRST invariant since $Q^2=0$).  This is sufficient to perform S-matrix calculations (with vertex operators of the type given for the superparticle above), but a naive application would require a measure that breaks manifest supersymmetry.  (For example, solving for the cohomology of the superparticle with this BRST operator in \cite{freeSBRST} required using the equivalent of the lightcone gauge.)  In principle, a covariant measure that avoids picture changing altogether (in particular, for the bosonic ghosts) can be found by methods similar to those used in \cite{Zwiebach}; we hope to return to this problem.  The cohomology of this BRST operator should also be checked:  The massless level follows from the previous analysis for the superparticle; the massive levels should follow from a similar lightcone analysis.

\section*{Acknowledgment}

This work was supported in part by the National Science Foundation Grant No.\ PHY-0354776.

\appendix

\mysection{Sp(2) components}

The matrix elements of the Sp(2) operators are
\begin{eqnarray}
\label{gammapls} \langle
p,q|\gamma^{\oplus}|r,s\rangle&=&C^{r,s}_{p,q}(
\sqrt{p}~\delta_{p,r+1}\delta_{q,s}~+~i\sqrt{q+1}~\delta_{p,r}\delta_{q+1,s})\\
\label{gammatildepls} \langle
p,q|\tilde{\gamma}^{\oplus}|r,s\rangle &=&C^{r,s}_{p,q}
(-i\sqrt{p}~\delta_{p,r+1}\delta_{q,s}~-~\sqrt{q+1}~\delta_{p,r}\delta_{q+1,s})
\end{eqnarray}
where
$C^{r,s}_{p,q}=i^{\frac{(r+s)(r+s+1)-(p+q)(p+q+1)}{2}}$. From these we can find ``inverse" operators, especially
\begin{equation}
\langle
p,q|\frac{1}{\tilde{\gamma}^{\oplus}}|r,s\rangle=~i^{r-p}C^{r,s}_{p,q}\sqrt{\frac{p!s!}{q!r!}}~
\delta_{q-p+r-s,1}~[\Theta_{p-q}\Theta_{s-q}~-~\Theta_{s-r}\Theta_{p-r}]
\end{equation}
It satisfies
\begin{equation}
\tilde{\gamma}^{\oplus}\frac{1}{\tilde{\gamma}^{\oplus}}\tilde{\gamma}^{\oplus}=\tilde{\gamma}^{\oplus},~~~~\tilde{\gamma}^{\oplus}\frac{1}{\tilde{\gamma}^{\oplus}}\rightarrow
I,~~~~\frac{1}{\tilde{\gamma}^{\oplus}}\tilde{\gamma}^{\oplus}\rightarrow
I
\end{equation}
The arrows means there is cancellation among the multiplied matrix
elements to the infinite ghost level. The subtle point of this
cancellation will be studied in Appendix B.

 Then we find that $\gamma^{\oplus}$ and
$\frac{1}{\tilde{\gamma}^{\oplus}}$ don't (anti)commute but give
\begin{eqnarray}
\label{ra}\langle
p,q|\{\frac{1}{\tilde{\gamma}^{\oplus}},\gamma^{\oplus}\}|r,s\rangle&=&i^{r-p+1}C^{r,s}_{p,q}\sqrt{\frac{p!s!}{q!r!}}~\delta_{q-p+r-s,0}
[(\Theta_{p-q}+\Theta_{p-q-1})(\Theta_{s-q}+\Theta_{s-q-1})\nonumber\\
&&~~~-(\Theta_{s-r-1}+\Theta_{s-r})(\Theta_{p-r-1}+\Theta_{p-r})]\\
\langle
p,q|[\frac{1}{\tilde{\gamma}^{\oplus}},\gamma^{\oplus}]|r,s\rangle&=&i^{r-p+1}C^{r,s}_{p,q}\sqrt{\frac{p!s!}{q!r!}}~\delta_{q-p+r-s,0}
[(\Theta_{p-q}-\Theta_{p-q-1})(\Theta_{s-q}+\Theta_{s-q-1})\nonumber\\
&&~~~-(\Theta_{s-r-1}-\Theta_{s-r})(\Theta_{p-r-1}+\Theta_{p-r})]\nonumber\\
&=&2i\delta_{p,q}\delta_{r,s}
\end{eqnarray}
Some interesting and useful commutators are
\begin{eqnarray}
\label{usefulcom} \langle
p,q|[\gamma^{\oplus},\{\frac{1}{\tilde{\gamma}^{\oplus}},\gamma^{\oplus}\}]|r,s\rangle
&=&-4i^{r-p}C^{r,s}_{p,q}(\sqrt{r+1}\delta_{p,q}\delta_{r+1,s}+\sqrt{p}\delta_{p,q+1}\delta_{r,s})\\
\langle
p,q|[\gamma^{\oplus}\gamma^{\oplus},\{\frac{1}{\tilde{\gamma}^{\oplus}},\gamma^{\oplus}\}]|r,s\rangle
&=&-8i^{r-p+1}C^{r,s}_{p,q}(\sqrt{(r+2)(r+1)}\delta_{p,q}\delta_{r+2,s}\nonumber\\
&+&\sqrt{p(p-1)}\delta_{p,q+2}\delta_{r,s}-2\sqrt{p(r+1)}\delta_{p,q+1}\delta_{r+1,s})\nonumber\\
\end{eqnarray}
\begin{eqnarray}
\langle
p,q|[\gamma^{\oplus}\gamma^{\oplus},\{\frac{1}{\tilde{\gamma}^{\oplus}},\gamma^{\oplus}\}]\gamma^{\oplus}|r,s\rangle
&=&16i^{r-p}C^{r,s}_{p,q}(\sqrt{(r+3)(r+2)(r+1)}\delta_{p,q}\delta_{r+3,s}\nonumber\\
&+&\sqrt{p(p-1)(r+1)}\delta_{p,q+2}\delta_{r+1,s}\nonumber\\
&-&2\sqrt{p(r+2)(r+1)}\delta_{p,q+1}\delta_{r+1,s})
\end{eqnarray}
Using (\ref{gammapls}),(\ref{gammatildepls}) and (\ref{ra}) we find
\begin{eqnarray}
\label{gammasqr}
\bar{\pi}\gamma^{\oplus}\gamma^{\oplus}\theta&=&\sum_{pq}[\sqrt{p(p-1)}~(-1)^{p+1}i^{p+q+1}~\bar{\pi}^{p,q}~\theta^{q,p-2}\nonumber\\
&&~~+2\sqrt{p(q+1)}~(-1)^{q+1}i^{p+q+1}~\bar{\pi}^{p,q}~\theta^{q+1,p-1}\nonumber\\
&&~~+\sqrt{(q+1)(q+2)}~(-1)^{p+1}i^{p+q+1}~\bar{\pi}^{p,q}~\theta^{q+2,p}]
\end{eqnarray}
\begin{equation}
\label{pisqr}
\f{1}{2}\bar{\pi}\tilde{\gamma}^{\oplus}\pi~=~\sum_{pq}\sqrt{p}~(-1)^{q+1}~\bar{\pi}^{p,q}~\pi^{q,p-1}~~~~~~~~~~~~~~
\end{equation}
\begin{eqnarray}
\label{gammadel}
\bar{\pi}\gamma^{\oplus}\gamma^{a}\theta&=&i\sum_{pq}[\sqrt{p}~(-1)^{p}~\bar{\pi}^{p,q}~\gamma^{a}~\theta^{q,p-1}~~~~~~~~~~~~~~~~~\nonumber\\
&&~~~~~~~+\sqrt{q+1}~(-1)^{q+1}~\bar{\pi}^{p,q}~\gamma^{a}~\theta^{q+1,p}]
\end{eqnarray}
\begin{eqnarray}
\label{R}
R^{a}&=&\f{1}{4}~\bar{\theta}\{\frac{1}{\tilde{\gamma}^{\oplus}},\gamma^{\oplus}\}\gamma^{a}\theta
\nonumber\\
&=&\f{1}{2}\sum_{pqr}(-1)^{(q-p+r+1)r-qp-p}i^{q-p+1}\sqrt{\frac{p!(q-p+r)!}{q!r!}}\bar{\theta}^{p,q}~\gamma^{a}~\theta^{q-p+r,r}\nonumber\\
&&~~~[(\Theta_{p-q}+\Theta_{p-q-1})(\Theta_{r-p}+\Theta_{r-p-1})\nonumber\\
&&~~~-(\Theta_{q-p-1}+\Theta_{q-p})(\Theta_{p-r-1}+\Theta_{p-r})]
\end{eqnarray}
\begin{eqnarray}
\label{G}
R^{\oplus}&=&\f{1}{4}~\bar{\theta}\{\frac{1}{\tilde{\gamma}^{\oplus}},\gamma^{\oplus}\}\gamma^{\oplus}\theta
\nonumber\\
&=&\f{1}{2}\sum_{pqr}(-1)^{(q-p+r+2)r-qp-p}\sqrt{\frac{p!(q-p+r+1)!}{q!r!}}\bar{\theta}^{p,q}~\theta^{q-p+r+1,r}\nonumber\\
&&~~~[(\Theta_{p-q}+\Theta_{p-q-1})(\Theta_{r-p+1}+2\Theta_{r-p}+\Theta_{r-p-1})\nonumber\\
&&~~~-(\Theta_{q-p-1}+\Theta_{q-p})(\Theta_{p-r-2}+2\Theta_{p-r-1}+\Theta_{p-r})]
\end{eqnarray}

The component fields defined above satisfy
\begin{equation}
\{\pi_{p,q},\theta^{r,s}]~=~\delta^{r}_{p}\delta^{s}_{q}
\end{equation}

\mysection{Subtle points in Sp(2) operators}

In this appendix we explain some subtle points about
$\frac{1}{\tilde{\gamma}^{\oplus}}$, due to the infinite dimensional structure of Sp(2) operators.

Consider the commutator
\begin{equation}
\{~\bar{\pi}\tilde{\gamma^{\oplus}}\theta~,~[~\stackrel{(1)}{\theta}\stackrel{(0)}{\theta}~,~R^{a}~]~\}
\end{equation}
where
\begin{eqnarray}
\stackrel{(0)}{\theta}&=&\langle0|e^{ia_{\oplus}a_{\ominus}}|\theta\rangle\nonumber\\
&=&\theta_{0}~+~\tilde{\theta}\nonumber\\
&=&\theta_{0}~+~\theta^{\oplus\ominus}~+~\theta^{\oplus\oplus\ominus\ominus}~+~\theta^{\oplus\oplus\oplus\ominus\ominus\ominus}~+~\cdots\nonumber\\
\stackrel{(1)}{\theta}&=&\langle0|e^{ia_{\oplus}a_{\ominus}}(ia_{\ominus})|\theta\rangle\nonumber\\
&=&2i(
\theta^{\oplus}~-~\sqrt{2}\theta^{\oplus\oplus\ominus}~+\sqrt{3}\theta^{\oplus\oplus\oplus\ominus\ominus}~-~\sqrt{4}\theta^{\oplus\oplus\oplus\oplus\ominus\ominus\ominus}~+~\cdots)
\end{eqnarray}
We will also need:

\noindent for $n\geq0$
\begin{eqnarray}
\stackrel{(n)}{\theta}&=&\langle0|e^{ia_{\oplus}a_{\ominus}}(ia_{\ominus})^{n}|\theta\rangle\nonumber\\
&=&\sum_{k=0}^{\infty}~(-1)^{k(n+k+1)}i^{\f{n(n+1)}{2}}(1+\Theta_{n-\f{1}{2}})\sqrt{\frac{(n+k)!}{k!}}\theta^{n+k,k}
\end{eqnarray}
\noindent for $n<0$
\begin{eqnarray}
\stackrel{(n)}{\theta}&=&\langle0|e^{ia_{\oplus}a_{\ominus}}(ia^{\dag\oplus})^{|n|}|\theta\rangle\nonumber\\
&=&\sum_{k=0}^{\infty}~(-1)^{k(|n|+k+1)}i^{\f{|n|(|n|+1)}{2}}(1+\Theta_{|n|-\f{1}{2}})\sqrt{\frac{k!}{(|n|+k)!}}\theta^{k,|n|+k}
\end{eqnarray}
The above double commutator should vanish because the inner
one involves only $\theta$. However, if we apply
the Jacobi identity we see
\begin{eqnarray}
\label{doublecomm}
\{\bar{\pi}\tilde{\gamma^{\oplus}}\pi,[\stackrel{(1)}{\theta}\stackrel{(0)}{\theta},R^{a}]\}&=&
0~=~[\{\bar{\pi}\tilde{\gamma^{\oplus}}\pi,\stackrel{(1)}{\theta}\stackrel{(0)}{\theta}\},R^{a}]
~-~\{\stackrel{(1)}{\theta}\stackrel{(0)}{\theta},[\bar{\pi}\tilde{\gamma^{\oplus}}\pi,R^{a}]\}\nonumber\\
&&~~~=~[\tilde{0},R^{a}]~-~2~\{\stackrel{(1)}{\theta}\stackrel{(0)}{\theta},\bar{\pi}\gamma^{\oplus}\gamma^{a}\theta\}\nonumber\\
&&~~~=~[\tilde{0},R^{a}]~+~2~\stackrel{(1)}{\theta}\gamma^{a}\stackrel{(1)}{\theta}+~4~\stackrel{(0)}{\theta}\gamma^{a}\stackrel{(2)}{\theta}
\end{eqnarray}
where we have introduced the scalar $\tilde 0$ defined by
\begin{equation*}
[\f{1}{2}\bar{\pi}\gamma^{\oplus}\pi,\stackrel{(n)}{\theta}]~~(n\geq-1)\ \equiv\ -i^{\f{n(n+1)}{2}+n}(1+\Theta_{|n|-\f{1}{2}})\sqrt{\frac{(n+1)!}{0!}}(\bar{\pi}^{n+1,0}-\bar{\pi}^{n+1,0})
\end{equation*}
\vskip-.3in
\begin{eqnarray}
&&+(-1)^{n}i^{\f{n(n+1)}{2}+n}(1+\Theta_{|n|-\f{1}{2}})\sqrt{\frac{(n+2)!}{1!}}(\bar{\pi}^{n+2,1}-\bar{\pi}^{n+2,1})\nonumber\\
&&+~\cdots\nonumber\\
&&+(-1)^{(k+2)(n+k+2)-1}i^{\f{n(n+1)}{2}+n}(1+\Theta_{|n|-\f{1}{2}})\sqrt{\frac{(n+k+1)!}{k!}}(\bar{\pi}^{n+k+1,k}-\bar{\pi}^{n+k+1,k})\nonumber\\
&&+~\cdots\nonumber\\
&&\equiv\tilde{0}\ \rightarrow\ 0
\end{eqnarray}
and
\begin{equation}
[\stackrel{(n)}{\theta},\bar{\pi}\gamma^{\oplus}\gamma^{a}
\theta]~=~-i^{n}(1+\Theta_{n-\f{1}{2}})\gamma^{a}\stackrel{(n+1)}{\theta}
\end{equation}
Unfortunately, $\tilde{0}$ is not zero in the presence of
$R^{a}$. Let's assume the collective $\on{(n)}\theta{}_k $ has
$k+1$ terms instead of an infinite number of terms.
Then $[\f{1}{2}\bar{\pi}\gamma^{\oplus}\pi,\on{(n)}\theta{}_k ]$
gives
$$ (-1)^{(k+2)(n+k+2)-1} i^{\f{n(n+1)}{2}+n}
(1+\Theta_{|n|-\f{1}{2}})\sqrt{\frac{(n+k+1)!}{k!}}\bar{\pi}^{n+k+1,k} $$
Finally, one can see that
$$ (-1)^{(k+2)(n+k+2)-1}
i^{\f{n(n+1)}{2}+n} (1+\Theta_{|n|-\f{1}{2}})
\sqrt{\frac{(n+k+1)!}{k!}} [\bar{\pi}^{n+k+1,k},R^{a} ] $$ gives
$~-i^{n}(1+\Theta_{|n|-\f{1}{2}})\on{(n+1)}\theta{}_k $. So if we
make the collective $\on{(n)}\theta{}_k $ have an infinite number
of terms by $k\rightarrow\infty$ then $\tilde{0}$ produces a
nonzero result in the commutator with $R^{a}$. This exactly
cancels the remaining terms in (\ref{doublecomm}) to make the
double commutator consistent.

Keeping this subtle point in mind let's consider the following
transformation
\begin{eqnarray}
Q'_{free}&=&e^{iR^{a}p_{a}}\tilde{Q}_{free}e^{-iR^{a}p_{a}}\nonumber\\
&=&e^{iR^{a}p_{a}}(\f{1}{2}c\,\square~-\f{1}{2}\bar{\pi}\gamma^{\oplus}\gamma^{\oplus}\theta
b~+~\tilde{0}b~+[-iR^{a}p_{a},-\f{1}{2}\bar{\pi}\gamma^{\oplus}\gamma^{\oplus}\theta
~+~\tilde{0}]b\nonumber\\
&&+\
\f{1}{4}\bar{\pi}\tilde{\gamma}^{\oplus}\pi~+~[-iR^{a}p_{a},\f{1}{4}\bar{\pi}\tilde{\gamma}^{\oplus}\pi]
~+~\f{1}{2}[-iR^{a}p_{a},[-iR^{a}p_{a},\f{1}{4}\bar{\pi}\tilde{\gamma}^{\oplus}\pi]]\nonumber\\
&&-\
\f{i}{2}\bar{\pi}\gamma^{\oplus}\rlap/p\theta~+~[-iR^{a}p_{a},-\f{i}{2}\bar{\pi}\gamma^{\oplus}\rlap/p\theta])e^{-iR^{a}p_{a}}
\end{eqnarray}
where $\tilde{0}b$ vanishes in $Q'_{free}$ but will give a
nontrivial contribution in the presence of $R^{a}$. We will
determine this term from nilpotency of $Q'_{free}$.

In terms of $R^{\oplus}$ and $\stackrel{(n)}{\theta}$,
$\tilde{Q}_{free}$ is
\begin{eqnarray}
\label{secondfreeq}
\tilde{Q}_{free}&=&\f{1}{2}c\,\square~-\f{1}{2}\bar{\pi}\gamma^{\oplus}\gamma^{\oplus}\theta
b~-\stackrel{(2)}{\theta}\gamma^{a}\stackrel{(0)}{\theta}p_{a}b~+~\f{1}{2}\stackrel{(1)}{\theta}\gamma^{a}\stackrel{(1)}{\theta}p_{a}b\nonumber\\
&&~~~~+~\tilde{0}b~+~[-iR^{a},\tilde{0}]p_{a}b~+~\f{1}{4}\bar{\pi}\tilde{\gamma}^{\oplus}\pi~-~\tilde{0}^{a}p_{a}\nonumber\\
&&~~~~+~\f{1}{2}[-iR^{a}p_{a},-\tilde{0}^{b}p_{b}]~+~\f{1}{2}R^{\oplus}\square
\end{eqnarray}
where
\begin{equation}
[-iR^{a}p_{a},-\f{1}{2}\bar{\pi}\gamma^{\oplus}\gamma^{\oplus}\theta]b
~=~-\stackrel{(2)}{\theta}\gamma^{a}\stackrel{(0)}{\theta}p_{a}b~+~\f{1}{2}\stackrel{(1)}{\theta}\gamma^{a}\stackrel{(1)}{\theta}p_{a}b
\end{equation}
and we have defined the vector $\tilde 0^a$
\begin{equation}
\label{zerotildea}
\tilde{0}^{a}\equiv-[-iR^{a},\f{1}{4}\bar{\pi}\tilde{\gamma}^{\oplus}\pi]+\f{i}{2}\bar{\pi}\gamma^{\oplus}\gamma^{a}\theta
\end{equation}
The nilpotency of $Q'_{free}$ implies
\begin{equation}
\label{nilpotency}
-\{-\tilde{0}^a,\f{1}{2}R^{\oplus}\}p_{a}\square\ =\
\f{1}{2}(-\stackrel{(2)}{\theta}\gamma^{a}\stackrel{(0)}{\theta}
+\f{1}{2}\stackrel{(1)}{\theta}\gamma^{a}\stackrel{(1)}{\theta}+[-i
R^{a},\tilde{0}])p_{a}\square
\end{equation}
\begin{eqnarray}
[\f{1}{2}\bar{\pi}\gamma^{\oplus}\gamma^{\oplus}\theta,\f{1}{2}[-iR^{a}p_{a},-\tilde{0}^{b}p_{b}]]~b
&=&-[-\f{1}{2}\bar{\pi}\gamma^{\oplus}\gamma^{\oplus}\theta~+~\tilde{0},\f{1}{2}R^{\oplus}]\square~b\nonumber\\
&=&~\f{1}{4}(3i\stackrel{(1)}{\theta}\stackrel{(2)}{\theta}-2i\stackrel{(0)}{\theta}\stackrel{(3)}{\theta}-2[\tilde{0},R^{\oplus}])\square~b\nonumber\\
\end{eqnarray}
 From (\ref{zerotildea}) we can find $\tilde{0}^{a}$ as
\begin{equation}
\tilde{0}^{a}~=~\f{i}{4}\on{(1)}\pi{}_\infty \gamma^{a}\stackrel{(0)}{\theta}-\f{i}{4}\stackrel{(0)}{\pi_{\infty}}\gamma^{a}\stackrel{(1)}{\theta}
\end{equation}
where
\begin{equation}
\stackrel{(n)}{\chi_{\infty}}\equiv
\lim_{k\rightarrow\infty}(-1)^{k(n+k+1)}(-i)^{\f{n(n+1)}{2}}\sqrt{\frac{(n+k+1)!}{k!}}\chi^{n+k,k}
\end{equation}
for $\chi=\theta,\pi$.  Inserting this into (\ref{nilpotency}) we find
\begin{eqnarray}
~[i
R^{a},\tilde{0}]&=&\stackrel{(1)}{\theta}\gamma^{a}\stackrel{(1)}{\theta}\\
~[\tilde{0},R^{\oplus}]&=&2i\stackrel{(1)}{\theta}\stackrel{(2)}{\theta}
\end{eqnarray}
One solution for $\tilde{0}$ is
\begin{equation}
\tilde{0}=\on{(1)}\pi{}_\infty \stackrel{(1)}{\theta}
\end{equation}

Now $\tilde{Q}_{free}$ is nilpotent, as it should be. However, the
origin of collective nonminimal fields is that $R^{a}$ produces
$\bar{\pi}\gamma^{\oplus}\gamma^{a}\theta$ using
$[\bar{\pi}\tilde{\gamma}^{\oplus}\pi,\stackrel{(n)}{\theta}]\rightarrow0$.
(The key feature of $R^{a}$ is $\stackrel{(n)}{\theta}$, see
(\ref{R}).)  And this implies that $R^{a}$ and $R^{\oplus}$
commute with $\bar{\pi}\gamma^{\oplus}\gamma^{\oplus}\theta$ only
up to these collective nonminimal fields (see (\ref{usefulcom})).
So these collective fields are just mathematical objects to
compensate terms like $[R^i ,\tilde{0}]$. Therefore, a physical
(but not mathematical) equivalent is to drop $\tilde{0}^{(a)}$ and
$\stackrel{(n)}{\theta}$  and regard $R^i $ as terms commuting
with $\bar{\pi}\gamma^{\oplus}\gamma^{\oplus}\theta$ in
(\ref{secondfreeq}). The resulting $\tilde{Q}_{free}$ is simply
\begin{equation}
\label{simplefreeQ}
\tilde{Q}_{free}=\f{1}{2}c\,\square~-\f{1}{2}\bar{\pi}\gamma^{\oplus}\gamma^{\oplus}\theta
b~+~\f{1}{4}\bar{\pi}\tilde{\gamma}^{\oplus}\pi~+~\f{1}{2}R^{\oplus}\square
\end{equation}
with
$[\bar{\pi}\gamma^{\oplus}\gamma^{\oplus}\theta,R^{\oplus}]\sim
0$.

Our results for
$\on\circ Q{}'_{YMB}$(\ref{ordiym}),~$\on\circ Q{}_{SYMB}$(\ref{sym})
and
 $Q''_{SYMB}$(\ref{arbsymsimple}) all reflect this prescription. $\tilde{\theta}$ in
 these BRST operators will produce only $\pi^{\oplus}$,~
 $\theta^{\oplus}$ and $\theta^{\oplus\oplus}b$ ($\theta^{\oplus\oplus}b$ will only appear in
 (\ref{ordiym})) dropping $\stackrel{(1)}{\theta}$ and
 $\stackrel{(2)}{\theta}b$.
 If we want to be rigorous $\stackrel{(n)}{\theta}$ and
 $\tilde{0}^{(a)}$ should be kept and the commutator
 $[R^i ,\bar{\pi}\gamma^{\oplus}\gamma^{\oplus}\theta]$
 should be calculated for both to compensate terms from
 $\tilde{0}^{(a)}$.

If we consider this mathematical rigor for
$Q'_{YMB}$(\ref{ordiym}) we find
\begin{eqnarray}
Q^{collective}_{YMB}&=&e^{iR^{a}\nabla_{a}}[~\f{1}{2}c(\square-\bar{\pi}\gamma^{ab}\theta
F_{ab})~+~\f{1}{2}R^{\oplus}(\square-\bar{\pi}\gamma^{ab}\theta
F_{ab})\nonumber\\
&+&~\f{1}{2D}[iR^{a},\tilde{0}_{a}](\square-\bar{\pi}\gamma^{ab}\theta
F_{ab})~-\f{1}{2}\bar{\pi}\gamma^{\oplus}\gamma^{\oplus}\theta
b~+~\f{1}{4}\bar{\pi}\tilde{\gamma}^{\oplus}\pi\nonumber\\
&+&\tilde{0}b~-~\stackrel{(2)}{\theta}\gamma^{a}\stackrel{(0)}{\theta}\nabla_{a}b~+~\f{1}{2}\stackrel{(1)}{\theta}\gamma^{a}\stackrel{(1)}{\theta}\nabla_{a}b~-~[iR^{a},\tilde{0}]\nabla_{a}b\nonumber\\
&-&~\tilde{0}^{a}\nabla_{a}\nonumber\\
&-&\f{1}{2}\{(cb-\f{1}{2})+R^{\oplus}b+\f{1}{2D}[iR^{a},\tilde{0}_{a}]b\}\nonumber\\
&&\times~\bar{\theta}\frac{1}{\tilde{\gamma}^{\oplus}}\gamma^{ab}\theta[F_{ab},\nabla_{c}](-\stackrel{(2)}{\theta}\gamma^{c}\stackrel{(0)}{\theta}~+~\f{1}{2}\stackrel{(1)}{\theta}\gamma^{c}\stackrel{(1)}{\theta}~-~[iR^{c},\tilde{0}])\nonumber\\
&+&\f{1}{2}\{c+R^{\oplus}+\f{1}{2D}[iR^{a},\tilde{0}_{a}]\}\nonumber\\
&&\times~\bar{\theta}\frac{1}{\tilde{\gamma}^{\oplus}}\gamma^{ab}\theta[F_{ab},\nabla_{c}]\tilde{0}^{c}~]~e^{-iR^{a}\nabla_{a}}~|~_{\mbox{\scriptsize
linear in $F$},~[\nabla,F]=0}
\end{eqnarray}
where
\begin{eqnarray}
D&=& \mbox{dimension of space-time}\quad (\mbox{10 here})\nonumber\\
\mbox{and}~~\square&=&-\nabla^{a}\nabla_{a}
\end{eqnarray}
If we drop $\stackrel{(n)}{\theta}$ and $\tilde{0}^{(a)}$ and
regard $[\gamma^{\oplus},\frac{1}{\tilde{\gamma}^{\oplus}}]=0$ (which
is equivalent to
$[R^i ,\bar{\pi}\gamma^{\oplus}\gamma^{\oplus}\theta]]=0$) we
come back to $Q'_{YMB}$(\ref{ordiym}). The last four lines do not
contribute to $Q'_{YMB}$ but are there for nonconstant
Yang-Mills background.

\mysection{Regularization}

In this appendix
we will consider a regularization procedure which will give the
prescription of the previous appendix. The motivation is the fact
that
$[\bar{\pi}\tilde{\gamma}^{\oplus}\pi,\stackrel{(n)}{\theta}]\rightarrow0$.
However, this does not exactly vanish, but leaves a piece of
$\on{(n+1)}\pi{}_\infty $. This remnant gives a nontrivial
contribution in the presence of
$\frac{1}{\tilde{\gamma}^{\oplus}}$, i.e.,
\begin{eqnarray}
\label{projection}
\langle0|e^{ia_{\oplus}a_{\ominus}}(ia_{\ominus})^{n}\tilde{\gamma}^{\oplus}|\theta\rangle&\rightarrow&0\nonumber\\
\langle0|e^{ia_{\oplus}a_{\ominus}}(ia_{\ominus})^{n}\tilde{\gamma}^{\oplus}\frac{1}{\tilde{\gamma}^{\oplus}}|\theta\rangle&\rightarrow&\stackrel{(n)}{\theta}
\end{eqnarray}

Now if we introduce some regularization parameter $z$ as
\begin{equation}
\langle p,q|\tilde{\gamma}^{\oplus}|r,s\rangle
\stackrel{regularized}{\longrightarrow} z^{p+r+s+r}\langle
p,q|\tilde{\gamma}^{\oplus}|r,s\rangle,~~~ z\rightarrow1
\end{equation}
then (\ref{projection}) becomes
\begin{eqnarray}
\langle0|e^{ia_{\oplus}a_{\ominus}}(ia_{\ominus})^{n}\tilde{\gamma}^{\oplus}|\theta\rangle_{k}&\rightarrow&z^{2n+2k+1}~\theta^{n+k+1,k}\nonumber\\
\langle0|e^{ia_{\oplus}a_{\ominus}}(ia_{\ominus})^{n}\tilde{\gamma}^{\oplus}\frac{1}{\tilde{\gamma}^{\oplus}}|\theta\rangle_{k}&\rightarrow&
z^{2n+2k+1}\on{(n)}\theta{}_k
\end{eqnarray}
where $k$ means the collective field has $k+1$ terms.
The regularization is to send $k$ to infinity with fixed $z$ $(<1)$.
Then the operator
$\tilde{\gamma}^{\oplus}\frac{1}{\tilde{\gamma}^{\oplus}}$ is a
projection operator which will project out the collective fields.
The free terms in any $Q_{backgroud}$ all have this projection
operator which goes to the identity in the absence of
$\frac{1}{\tilde{\gamma}^{\oplus}}$.
\begin{eqnarray}
-\f{1}{2}\bar{\pi}\gamma^{\oplus}\gamma^{\oplus}\theta b
~+~\f{1}{4}\bar{\pi}\tilde{\gamma}^{\oplus}\pi&-&i\f{1}{2}\nabla_{a}\bar{\pi}\gamma^{\oplus}\gamma^{a}\theta\nonumber\\
&\rightarrow&
-\f{1}{2}\bar{\pi}\gamma^{\oplus}\gamma^{\oplus}\Pi\theta b
~+~\f{1}{4}\bar{\pi}\tilde{\gamma}^{\oplus}\Pi\pi~-i\f{1}{2}\nabla_{a}\bar{\pi}\gamma^{\oplus}\gamma^{a}\Pi\theta~~~
\end{eqnarray}
The arrow means inserting the projection operator and dropping
collective fields after expansion of exponential factors.

If a collective field is truncated, i.e., with incomplete
beginning or ending components, then we cannot remove it
by this regularization procedure, but we can still avoid its
contribution. This situation occurs when we consider the
commutator
\begin{equation}
\{\eta\pi^{p,p+1},[\nabla_{0}\tilde{\theta},\bar{\pi}\gamma^{\oplus}\gamma^{a}\theta]\}=\{\eta\pi^{p,p+1},-i\nabla_{0}\gamma^{a}\theta^{\oplus}\}=0~~~(p>0)
\end{equation}
where $\eta$ is a constant fermionic field.  This implies
\begin{equation}
[\nabla_{0}\tilde{\theta},(\pi^{p,p}+\pi^{p+1,p+1})\eta]=0
\end{equation}
The second argument in the commutator is an example of a truncated
collective field. Actually, this commutator indeed vanishes if we
consider $\on{(1)}\theta$, which we drop in regularization. So for
consistent regularization we take this as vanishing.

This fact is applied for closure of the algebra
\begin{equation}
[\eta\theta^{q,q+1},\{~[\eta'\pi^{p,p+1},Q_{R}],Q_{R}\}]
\end{equation}
where $Q_{R}$ is any version of the BRST operator including
$R^i $. After canceling ghost-number-nonzero components this
commutator reduces to
\begin{equation}
~[\eta' C^{-}_{p},\eta C^{+}_{q}]
\end{equation}
where $C^{\pm}_{p}$ will be superstring constraints if we use
$Q_{sstring}$.
But the original commutator is just
\begin{equation}
\f{1}{2}[\eta\theta^{q,q+1},~[\eta'\pi^{p,p+1},\{Q_{R},Q_{R}\}]]
\end{equation}
and it is just zero due to nilpotency of $Q_{R}$. However, $\eta'
C^{-}_{p}$ has a term like $\eta\gamma^{a}(\pi^{p,p}+\pi^{p+1,p+1})$, which comes from
$[\eta'\pi^{p,p+1},\bar{\pi}\gamma^{\oplus}\gamma^{a}\theta]$. But
this combination of $\pi^{p,p}$ is just an example of truncated
collective fields. This truncated collective field will interact
with $\tilde{\theta}$ in $\eta C^{+}_{q}$ giving a nonzero
contribution in this apparently vanishing commutator.
This should be canceled by a
$[\eta'\pi^{p,p+1},\stackrel{(1)}{\theta}]$ contribution, which
we projected out by regularization. What this means is that for
consistent regularization we should take the commutator with
truncated collective fields,~$\tilde{\theta}$,~$R^i $ as
vanishing. For example, we should take
$[(\pi^{p,p}+\pi^{p+1,p+1}),\tilde{\theta}]$ as zero but we should
calculate
$\{(\theta^{p,p}+\theta^{p+1,p+1}),(\pi^{p,p}+\pi^{p+1,p+1})\}$ in
$[\eta' C^{-}_{p},\eta C^{+}_{p}]$, both of which come from
$\bar{\pi}\gamma^{\oplus}\gamma^{a}\theta$.

\mysection{Closure of constraints}

First of all, if one directly calculates $[\pi^{p,p+1},R^{\oplus}]$ one gets
\begin{eqnarray}
[\pi^{p,p+1},R^{\oplus}]&=&-\f34\sum_{r}(-1)^p\sqrt{p+1}\theta^{r,r}(\Theta_{r-p}+2\Theta_{r-p-1}+\Theta_{r-p-2})\nonumber\\
&&+\f{1}{4}\sum_{r}(-1)^{p}\sqrt{p+1}\theta^{r,r}[(\Theta_{p-r+1}+2\Theta_{p-r}+\Theta_{p-r-1})\nonumber\\
&\equiv&-\f34A+\f14B\nonumber\\
&=&-\f12(A-B)-\f{1}{4}(A+B)
\end{eqnarray}
But $A+B$ is just $(-1)^{p}\sqrt{p+1}\stackrel{(0)}{\theta}$, and
it will vanish when it acts on the projection operator $\Pi$. Also,
\begin{equation*}
A-B\ =\ -2(-1)^{p}\sqrt{p+1}\vartheta_p
\end{equation*}
in terms of $\vartheta$ (\ref{vartheta}).  When we calculate closure of the constraints, there are two types of
terms related to the above expression, i.e.,
\begin{eqnarray*}
~~~~~~~~~~\{[\eta\pi^{p,p+1},R^{\oplus}],[\eta'\theta^{q,q+1},\bar{\pi}\tilde{\gamma}^{\oplus}\pi]\}
\end{eqnarray*}
and
\begin{eqnarray*}
&&\{[\eta\pi^{p,p+1},R^{\oplus}],[\eta'\pi^{q,q+1},\bar{\pi}\gamma^{\oplus}\gamma^{a}\theta]\}\\
&&\{[\eta\pi^{p,p+1},\bar{\pi}\gamma^{\oplus}\gamma^{a}\theta],[\eta'\pi^{q,q+1},R^{\oplus}]\}
\end{eqnarray*}
where $\eta$ and $\eta'$ are constant spinors.

 The first type is always zero due to the symmetry of $\gamma^{\oplus}$ and
$\tilde{\gamma}^{\oplus}$. If $p\neq q$, $p\neq q+1$ and $p+1\neq
q$ the second type cancels because of the $A-B$ sign difference.
If $p=q$ we can express $[\eta\pi^{p,p+1},R^{\oplus}]$ as just
$-\sqrt{p+1}\eta(\theta^{p,p}-\theta^{p+1,p+1})$. If $p=q+1$ or
$p+1=q$ we can use
$\sqrt{p+1}\eta(\theta^{p,p}-\theta^{p+1,p+1})$. This becomes
clearer in a simpler situation,
\begin{equation*}
\{R^{\oplus},\bar{\pi}\gamma^{\oplus}\gamma^{a}\theta\}=0
\end{equation*}
Then we have
\begin{equation*}
\{\eta'\pi^{q,q+1},[\eta\pi^{p,p+1},\{R^{\oplus},\bar{\pi}\gamma^{\oplus}\gamma^{a}\theta\}]\}=0
\end{equation*}
But this implies
\begin{equation*}
[\{\eta\pi^{p,p+1},R^{\oplus}\},\{\eta'\pi^{q,q+1},\bar{\pi}\gamma^{\oplus}\gamma^{a}\theta\}]+[\{\eta'\pi^{q,q+1},R^{\oplus}\},\{\eta\pi^{p,p+1},\bar{\pi}\gamma^{\oplus}\gamma^{a}\theta\}]=0
\end{equation*}
This is one part of the closure of constraints.
A similar analysis shows
\begin{eqnarray}
&&[~\eta\hat{P}_{a}\gamma^{a}(\theta^{p,p}+\theta^{p+1,p+1}),~\eta'\hat{P}_{a}\gamma^{a}(\theta^{p,p}+\theta^{p+1,p+1})~]\nonumber\\
&\sim&[~\eta(\pi^{p,p}-\pi^{p+1,p+1}),~\f{i}{4}\tilde{R}_{a}~]~\eta'\gamma^{a}(\theta^{p,p}+\theta^{p+1,p+1})\nonumber\\
&&+~\eta\gamma^{a}(\theta^{p,p}+\theta^{p+1,p+1})~[~\f{i}{4}\tilde{R}_{a},\eta'(\pi^{p,p}-\pi^{p+1,p+1})~]
\end{eqnarray}
where $\sim$ means we should drop truncated collective fields.

Secondly, from the fact that
$[\tilde{\theta},\bar{\pi}\gamma^{\oplus}\gamma^{a}\theta]$=$\stackrel{(1)}{\theta}-i~\theta^{\oplus}$
$\sim$ $-i~\theta^{\oplus}$ due to regularization when
$\tilde{\theta}$ acts on terms from
$\bar{\pi}\gamma^{\oplus}\gamma^{a}\theta$, we should change the sign
of $\tilde{\theta}$. ( More precisely, they are all zero except for
$\pi^{1,1}$ from $\bar{\pi}\gamma^{\oplus}\gamma^{a}\theta$. For
only this term we can see this sign-change effect as explained in
the previous appendix.) This fact was implicitly expressed with the
projection operator $\Pi$ in the constraints.

Finally, one should be cautious about
$\{\bar{\pi}\gamma^{\oplus}\gamma^{a}\theta$ $,$
$\bar{\pi}\gamma^{\oplus}\gamma^{b}\theta \}$ $=$ $(-1)(-2)g^{ab}$
$\bar{\pi}\gamma^{\oplus}\gamma^{\oplus}\theta$. The ``$-$" sign
comes from the fact that OSp(2) gamma matrices (and therefore
$a^{\dag}$ and $a$) anticommute with ordinary gamma matrices. This
gives an additional sign when one calculates terms like
$[\eta\gamma^{a}\theta^{p,p}$ $,$ $\eta'\gamma^{b}\pi^{p,p}]$.

\mysection{ZJBV form of BRST}

The ZJBV form of the BRST operator follows from the
Hamiltonian form of the BRST operator
\begin{equation}
Q^{ZJBV}=\f{i}{2}\dot{\phi}^{A}\phi_{A}-\check{\phi}_{A}\{\phi^{A},Q^{H}]
\end{equation}
Then in our case,
\begin{equation}
Q^{ZJBV}_{sstring}=-\dot{X}\cdot
P^{0}~+\sum_{\pm}Q^{(\pm)ZJBV}_{sstring}
\end{equation}
where, e.g., for the $(+)$ term (for $(-)$, just add a $-$ for each ${}'$)
\begin{eqnarray*}
Q^{ZJBV}_{sstring}&=&\dot{c}b~+~\dot{\theta}\pi\nonumber\\
&+&~(\check{X}_{a}-\check{P}'_{a})\left[\left(c+\f{1}{2}\bar{\theta}\tilde{\gamma}^{\oplus}\theta\right)\hat{P}^{a}~-~\f{i}{2}\bar{\pi}\gamma^{\oplus}\gamma^{a}\theta\right.\\
&&-\left(i\theta_{0}\gamma_{a}\theta'_{0}~+2i\tilde{\theta}\gamma_{a}\theta'_{0}~+~i\tilde{\theta}\gamma_{a}\tilde{\theta}'~-\f{1}{2}~R'_{a}|_{>}\right)R^{\oplus}+~i\f{1}{2}\bar{\theta}^{\oplus}\theta_{0}~\theta_{0}\gamma^{a}\theta'_{0}\\
&&-\f{1}{2}\bar{\theta}^{\oplus}\gamma^{b}\gamma^{a}\theta_{0}\left(i\f{1}{2}\theta_{0}\gamma_{b}\theta'_{0}~+~2i\tilde{\theta}\gamma_{b}\theta'_{0}~+~i\tilde{\theta}\gamma_{b}\tilde{\theta}'-\f{1}{2}~R'_{a}|_{>}\right)\\
&&-\left.\f{1}{3}\bar{\theta}^{\oplus}\gamma^{a}\gamma^{b}\tilde{\theta}\left(2i\tilde{\theta}\gamma_{a}\theta'_{0}~+~i\f{5}{4}~\tilde{\theta}\gamma_{a}\tilde{\theta}'-\f{3}{4}~R'_{a}|_{>}\right)\right]~~~~~~~~~~~~\\
&-&~\check{c}\left[icc'+2~\bar{\pi}a^{\dagger\oplus}a^{\oplus}\theta|_{>}\right]\\
&-&~\check{b}\left(~-\f{1}{2}\hat{P}^{2}~-2ic'b~-icb'~-~i\bar{\theta}'\pi~-~i[\bar{\theta}(a^{\dag\oplus}a_{\oplus}-a^{\dag\ominus}a_{\ominus})\pi]'\right)\\
&+&\sum_{p>1}~\check{\theta}_{p,p+1}\left[\left\{i\left(c+R^{\oplus}+\f{1}{2}\bar{\theta}\tilde{\gamma}^{\oplus}\theta~\right)|_{>}\theta'^{p,p+1}\right\}\right.\\
&&-~2i[(p+1)-\sqrt{p+1}\frac{\sqrt{p}+\sqrt{p+2}}{2}]\Pi\theta^{p+1,p}b\nonumber\\
&&+(-1)^{p+1}i\f{1}{2}\sqrt{p+1}\Pi(\pi^{p,p}-\pi^{p+1,p+1})\\
&&\left.+(-1)^{p+1}i\f{1}{2}\sqrt{p+1}\gamma^{a}\Pi(\theta^{p,p}+\theta^{p+1,p+1})\mathcal{P}_{a}\right]\\
&-&\f{i}{2}~\check{\theta}_{0,1}\left\{\pi_{0}+(\gamma^{a}\theta_{0})\hat{P}_{a}+i\f{1}{2}(\gamma^{a}\theta_{0})\theta_{0}\gamma_{a}\theta'_{0}~\right.\\
&&\left.+\f{2}{3}\gamma^{a}\tilde{\theta}\left(2i\tilde{\theta}\gamma_{a}\theta'_{0}~+~i\f{5}{4}~\tilde{\theta}\gamma_{a}\tilde{\theta}'-\f{3}{4}~R'_{a}|_{>}\right)-\pi^{1,1}+\gamma^{a}\Pi\theta^{1,1}\mathcal{P}_{a}\right\}\\
&-&\sum_{p>1}~\check{\pi}_{p,p+1}\left[\left\{i\left(c+R^{\oplus}+\f{1}{2}\bar{\theta}\tilde{\gamma}^{\oplus}\theta~\right)|_{>}\pi^{p,p+1}\right\}'\right.\\
&&+(-1)^{p+1}i\sqrt{p+1}\vartheta^{p}
\left\{-\f{1}{2}\mathcal{P}^{2}+\f{1}{2}\hat{P}^{2}+i\bar{\theta}'\pi+i[\bar{\theta}(a^{\dag\oplus}a_{\oplus}-a^{\dag\ominus}a_{\ominus})\pi]'\right\}\\
&&+\f{1}{2}R^{\oplus}|_{>}\sum_{r}(-1)^{r^{2}+p+1}\sqrt{\frac{p+1}{r}}\gamma^{a}\theta'^{r-1,r}(\Theta_{r-p-1}+\Theta_{r-p-2})\mathcal{P}_{a}\\
&&-~(-1)^{p+1}i\sqrt{p+1}(\theta^{p,p}-\theta^{p+1,p+1})\\
&&\quad\times\left(~\f{1}{2}\hat{P}^{2}~+~i\bar{\theta}'\pi~+~i[\bar{\theta}(a^{\dag\oplus}a_{\oplus}-a^{\dag\ominus}a_{\ominus})\pi]'\right)\\
&&+~2i[(p+1)-\sqrt{p+1}\frac{\sqrt{p}+\sqrt{p+2}}{2}]\pi^{p+1,p}b\\
&&-(-1)^{p+1}i\f{1}{2}\sqrt{p+1}\gamma^{a}\Pi(\pi^{p,p}+\pi^{p+1,p+1})\mathcal{P}_{a}\\
&&+\f{1}{4}\sum_{r}(-1)^{r^{2}+p+1}\sqrt{\frac{p+1}{r}}\gamma^{a}\theta'^{r-1,r}(\Theta_{r-p-1}+\Theta_{r-p-2})\\
&&\ \ \times\left.\left(-\bar{\pi}\gamma^{\oplus}\gamma^{a}\theta|_{>}+i\bar{\theta}^{\oplus}\gamma^{a}\left(\pi_{0}+(\gamma^{b}\theta_{0})\hat{P}_{b}+i\f{1}{2}(\gamma^{b}\theta_{0})\theta_{0}\gamma_{b}\theta'_{0}~\right)+i\bar{q}^{\oplus}\gamma_{a}\tilde{\theta}\right)\right]
\end{eqnarray*}
\begin{eqnarray}
&-&\f{i}{2}~\check{\pi}_{0,1}\left\{\gamma^{a}\left(\pi_{0}+(\gamma^{b}\theta_{0})\hat{P}_{b}+i\f{1}{2}(\gamma^{b}\theta_{0})\theta_{0}\gamma_{b}\theta'_{0}~\right)\right.~~~~~~~~~~~~~~~~~~~~~~~~~~\nonumber\\
&&\quad\times\left(\hat{P}_{a}~+i\theta_{0}\gamma_{a}\theta'_{0}~+2i\tilde{\theta}\gamma_{a}\theta'_{0}~+~i\tilde{\theta}\gamma_{a}\tilde{\theta}'-\f{1}{2}~R'_{a}|_{>}~\right)\nonumber\\
&&+\f{2}{3}\gamma^{b}\gamma^{a}\tilde{\theta}\left(2i\tilde{\theta}\gamma_{a}\theta'_{0}~+~i\f{5}{4}~\tilde{\theta}\gamma_{a}\tilde{\theta}'-\f{3}{4}~R'_{a}|_{>}\right)\nonumber\\
&&\left.\quad\times\left(\hat{P}_{b}~+i\theta_{0}\gamma_{b}\theta'_{0}~+2i\tilde{\theta}\gamma_{b}\theta'_{0}~+~i\tilde{\theta}\gamma_{b}\tilde{\theta}'-\f{1}{2}~R'_{b}|_{>}~\right)\right.\nonumber\\
&&-\gamma^{a}\Pi\pi^{1,1}\left(\hat{P}_{a}~+i\theta_{0}\gamma_{a}\theta'_{0}~+2i\tilde{\theta}\gamma_{a}\theta'_{0}~+~i\tilde{\theta}\gamma_{a}\tilde{\theta}'-\f{1}{2}~R'_{a}|_{>}~\right)\nonumber\\
&&+2\theta^{1,1}\left(~-\f{1}{2}\hat{P}^{2}~-~i\bar{\theta}'\pi~-~i[\bar{\theta}(a^{\dag\oplus}a_{\oplus}-a^{\dag\ominus}a_{\ominus})\pi]'\right)\nonumber\\
&&+2\vartheta^{0}\times
\left\{-\f{1}{2}\left(\hat{P}_{a}~+i\theta_{0}\gamma_{a}\theta'_{0}~+2i\tilde{\theta}\gamma_{a}\theta'_{0}~+~i\tilde{\theta}\gamma_{a}\tilde{\theta}'~-\f{1}{2}~R'_{a}|_{>}\right)^{2}\right.\nonumber\\
&&\left.\quad+\f{1}{2}\hat{P}^{2}~+~i\bar{\theta}'\pi~+~i[\bar{\theta}(a^{\dag\oplus}a_{\oplus}-a^{\dag\ominus}a_{\ominus})\pi]'\right\}\nonumber\\
&&+\f{1}{2}R^{\oplus}|_{>}\sum_{r}(-1)^{r^{2}+1}\frac{1}{\sqrt{r}}\gamma^{a}\theta'^{r-1,r}(\Theta_{r-1}+\Theta_{r-2})\mathcal{P}_{a}\nonumber\\
&&-\f{i}{4}\sum_{r}(-1)^{r^{2}+1}\f{1}{\sqrt{r}}\gamma^{a}\theta'^{r-1,r}(\Theta_{r-1}+\Theta_{r-2})\nonumber\\
&&\quad\times\left.\left(-\bar{\pi}\gamma^{\oplus}\gamma^{a}\theta|_{>}+i\bar{\theta}^{\oplus}\gamma^{a}\left(\pi_{0}+(\gamma^{b}\theta_{0})\hat{P}_{b}+i\f{1}{2}(\gamma^{b}\theta_{0})\theta_{0}\gamma_{b}\theta'_{0}~\right)+i\bar{q}^{\oplus}\gamma_{a}\tilde{\theta}\right)\right\}\nonumber\\
&-&\sum_{q\neq
p+1}\check{\theta}_{p,q}[\theta^{p,q},Q_{sstring}\}~~~~~~~~~~~~~~~~~~~~~~~~~~~~~~~~~~~~~~~~~~~~~~~~~~~~~~~~~\nonumber\\
&-&\sum_{q\neq
p+1}\check{\pi}_{p,q}[\pi^{p,q},Q_{sstring}\}~~~~~~~~~~~~~~~~~~~~~~~~~~~~~~~~~~~~~~~~~~~~~~~~~~~~~~~~~~~
\end{eqnarray}
where \begin{equation} \Theta_{x}=\left\{\begin{matrix} 1&~x\geq0\\
0&~x<0
\end{matrix}\right.
\end{equation}
\setlength{\baselineskip}{15pt}

\end{document}